\documentclass[%
 reprint,
 amsmath,amssymb,
 aps,
]{revtex4-2}

\usepackage{graphicx}
\usepackage{dcolumn}
\usepackage{bm}

\begin{document}
\raggedbottom
\preprint{APS/123-QED}

\title{Topology and Kinetic Pathways of Colloidosome
Assembly and Disassembly}

\author{Raymond Adkins\textsuperscript{1,2},
Joanna Robaszewski\textsuperscript{1,3},
Seungwoo Shin\textsuperscript{1},
Fridtjof Brauns\textsuperscript{4},
Leroy Jia\textsuperscript{5},
Ayantika Khanra\textsuperscript{6},
Prerna Sharma\textsuperscript{6,7},
Robert Pelcovits\textsuperscript{8},
Thomas R. Powers\textsuperscript{8,9}
Zvonimir Dogic\textsuperscript{1,*}}

\email{Email for correspondence: zdogic@ucsb.edu}

\affiliation{$^1$Department of Physics, University of California, Santa Barbara, Santa Barbara, California 93106, USA}
\affiliation{$^2$Molecular Biophysics and Biochemistry, Yale University, New Haven, Connecticut 06510, USA}
\affiliation{$^3$Environmental Laboratory, US Army Engineer Research and Development Center, Concord, Massachusets 01742, USA}
\affiliation{$^4$Kavli Institute for Theoretical Physics, University of California, Santa Barbara, Santa Barbara, California 93106, USA}
\affiliation{$^5$Applied and Computational Mathematics Division, National Institute of Standards and Technology, Gaithersburg, Maryland 20899, USA}
\affiliation{$^6$Department of Physics, Indian Institute of Science, Bangalore 560012, India}
\affiliation{$^7$Department of Bioengineering, Indian Institute of Science, Bangalore 560012, India}
\affiliation{$^8$Department of Physics, Brown University, Providence, Rhode Island 02906, USA}
\affiliation{$^9$School of Engineering, Brown University, Providence, Rhode Island 02906, USA}

\date{\today}

\begin{abstract}
Liquid shells, such as lipid vesicles and soap bubbles, are ubiquitous throughout biology, engineered matter, and everyday life. Their creation and disintegration are defined by a singularity that separates a topologically distinct extended liquid film from a boundary-free closed shell. Such topology-changing processes are essential for cellular transport and drug delivery. However, their studies are challenging because of the rapid dynamics and small length scale of conventional lipid vesicles. We develop fluid colloidosomes, micron-sized analogs of lipid vesicles. We study their stability close to their disk-to-sphere topological transition. Intrinsic colloidal length and time scales slow down the dynamics to reveal vesicle conformations in real time during their assembly and disassembly. Remarkably, the lowest-energy pathway by which a closed vesicle transforms into a disk involves a topologically distinct cylinder-like intermediate. These results reveal universal aspects of topological changes in all liquid shells and a robust platform for the encapsulation, transport, and delivery of nanosized cargoes.
\end{abstract}


\maketitle

\section{Introduction}

Controlling the shape and topology of thin elastic sheets is essential for creating reconfigurable and adaptable materials. For inspiration, one can turn toward living matter, where shape morphing enables diverse life-sustaining processes. On macroscopic scales, plants grow sheet-like tissues into intricate leaves and flowers that reconfigure in response to sunlight~\cite{Nath2003,Liang2011}. At the mesoscopic scale, micron-thick epithelial sheets undergo highly choreographed morphological and topological transitions to assume complex shapes that define three-dimensional organs~\cite{Metzger2008, Mitchell2022}. At subcellular scales, nanometer-thick lipid bilayers are a distinct category of thin sheets that lack in-plane shear modulus but form complex shapes and topologies~\cite{kozlovsky2002stalk,conner2003regulated,jahn2003membrane,nixon2016increased,shin2018visualization}. Translating these structures and motifs into the realm of synthetic materials represents a challenge and an opportunity. For example, patterning in-plane strains into mesoscale stimuli-responsive solid-elastomeric sheets generated designable and controllable three-dimensional shape morphing materials~\cite{Klein2007,armon2011geometry,kim2012designing,sydney2016biomimetic}. On microscopic scales, lipid vesicles inspired the creation of synthetic analogues with potential applications in transport, encapsulation, and drug delivery~\cite{evans1990entropy,seifert1997configurations,Discher1999,Dinsmore2002,baumgart2003imaging,xu2021transmembrane}. In comparison to solid sheets controlling the morphology of fluid membranes is significantly more challenging. 

We seek to control the morphology and topology of fluid membranes by balancing the bending and edge energy. The edge energy of a flat disk increases with its size while the bending energy does not depend on the size of edgeless spherical vesicles~\cite{Umeda2005,noguchi2006dynamics,reynwar2007aggregation,Hu2012}. Thus, with increasing size the edge energy destabilizes a flat disk, inducing a mechanical instability that generates edgeless vesicles. To observe and quantify such transitions we develop fluid colloidosomes, which are colloidal analogs of lipid vesicles, assembled from monodisperse sub-micron-sized rod-like particles. By modulating colloidosome size {\it in situ} we balance edge and bending energy to control transitions between vesicle-like spheres and disks in real-time. In particular, we elucidate the complex vesicle disassembly pathways involving intermediate states, whose topologies are more complex than both the initial and the final state. The rich energy landscape associated with the disk-to-vesicle transition provides a unique platform to control the morphology and topology of fluid membranes. Colloidosomes combine desirable features of fluid bilayers, including reconfigurability, self-healing, and topological transitions with those of solid elastic sheets, such as programmability and control. They also provide insight into universal assembly and disassembly pathways that apply to all vesicle-like materials.

\section{Results}

\subsection{Colloidosomes assume energy-minimizing shapes}

In the presence of nonadsorbing polymers, charged rod-shaped particles experience attractive depletion interactions that favor their lateral association~\cite{asakura1954interaction}. Such effective interactions can drive the assembly of one-rod-length-thick fluid monolayer membranes, which assume a flat disk-like shape~\cite{Barry2010,yang2012self,kang2016entropic}. Similarly to lipid bilayers ~\cite{Helfrich1973,faizi2020fluctuation}, colloidal monolayers are described by the Helfrich free energy:

\begin{equation}\label{eq:Helfrich}
    E = \int \Big[ \frac{\kappa}{2} (2H)^2 +  \Bar{\kappa}K \Big] dA + \gamma \int dL, 
\end{equation}

where $H$ and $K$ are the mean and Gaussian curvatures, $\kappa$ and $\Bar{\kappa}$ are the associated moduli, $\gamma$ is the edge tension, and $dA$ and $dL$ are respectively the surface area and boundary elements. It follows, that the energy of a flat disk-shaped membrane is $E_\mathrm{mem} = 2 \gamma \sqrt{\pi A}$, where $A$ is disk area, while the edgeless spherical vesicle energy is $E_\mathrm{vesicle} = 4 \pi (2 \kappa + \Bar{\kappa})$. Vesicle bending energy is size-independent, while disk energy increases with its size. Therefore, beyond $A_1^*= 4 \pi (2 \kappa + \Bar{\kappa})^2 / \gamma^2 $ flat disks are metastable with respect to closed vesicles. Once $A > A_2^* = 4 A_1^*$, the disk-to-vesicle transition barrier disappears, and flat disks are unstable. These considerations explain the relative stability of lipid-bilayers closed vesicles 2D disk-shaped  colloidal membranes~\cite{helfrich1974size,fromherz1983lipid}. Both systems have comparable edge energy of a few hundred $k_\text{B}T$/µm~\cite{portet2010new} (Fig. S1), but widely-different bending moduli. General arguments predicts that $\kappa \propto (\text{thickness})^n$, where $n \geq 2.5$~\cite{Szleifer1988,rawicz2000effect}. Therefore micron-thick membranes will have orders of magnitude larger bending rigidity when compared to nanometer-thick lipid bilayers~\cite{Balchunas2019}. 

\begin{figure*}[t]
\centering
\includegraphics[width=0.65\textwidth]{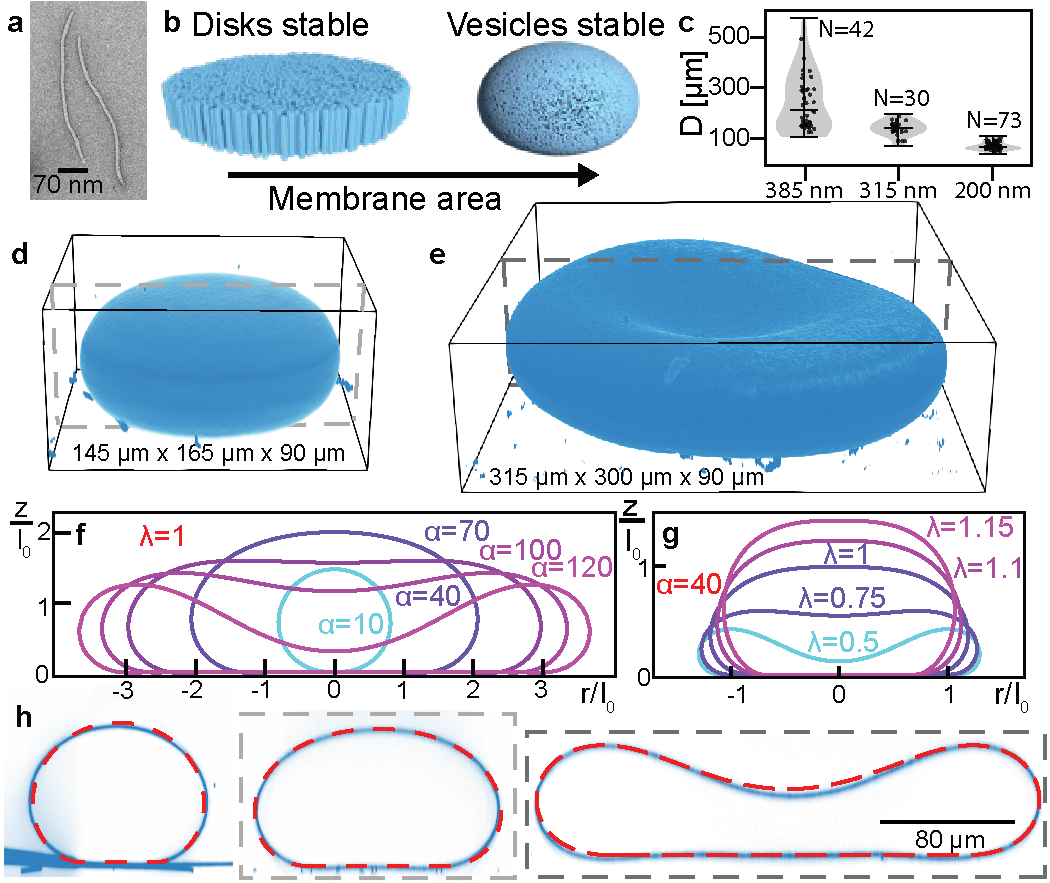}
\caption{\textbf{Colloidosomes minimize membrane elastic energy.} \textbf{a,} Electron microscopy image of rod-like nano385. \textbf{b,} Fluid membranes transition from a flat disk to edgeless vesicles with increasing area.  \textbf{c,} Vesicle diameter, $D$, decreases with decreasing virus length. \textbf{d-e,} Smaller colloidosomes have rounded shapes while larger ones sag under gravity. \textbf{f,} Predicted colloidosome shapes with varying $\alpha$ and fixed $\lambda$ = 1. \textbf{g,} Colloidosome shapes with varying inflation parameter $\lambda$ at fixed $\alpha$ = 40. \textbf{h,} Cross sections of three colloidosomes using measured values of $\kappa$, $V$ and $A$. Dashed lines show the energy-minimizing contour. }
	\label{fig:Fig1}
\end{figure*}

To assemble colloidosomes, we reduced $A_1^*$. Instead of previously used micron-long rods~\cite{Barry2010}, we assembled membranes from 385 nm long virus-like rods (nano385), which decreased $\kappa$. Simultaneously, to assemble membranes larger than $A_1^*$ we improved the purity of critical components which yielded larger structures (Materials and Methods). Upon mixing nano385 and Dextran, small disk-shaped membranes formed throughout the sample. Such membranes coalesced laterally \cite{Zakhary2014}, growing in size while sedimenting towards the chamber bottom. Once at the bottom, now-large membranes formed curved sheets (Fig. S4). Reasoning that gravity suppressed vesicle closure, we inverted the sample and waited for an additional 18 hours. At that point, we observed large unilamellar and multi-lamellar vesicle-like colloidal membranes or colloidosomes (Movie S1). Smaller colloidosomes were nearly spherical (Fig. \ref{fig:Fig1}d), while larger ones were shaped as biconcave disks (Fig. \ref{fig:Fig1}e).  Isolated axisymmetric colloidosomes are described by their cross-sectional contours (Fig. S3). In a field of view, we observed hundreds of partially or fully closed structures (Movie S1).  

The above-described model predicts that decreasing rod contour length decreases $\kappa$ which in turn should reduce $A_1^*$ and the average colloidosome size. To test this effect, we assembled colloidosomes from 385, 315 and 200 nm long particles. We found that the mean and minimum size decreased with decreasing filament length, indicating tunable colloidosome size (Fig. \ref{fig:Fig1}c).

To explain colloidosome shapes we modified the Helfrich energy to account for gravity and a Lagrange multiplier enforcing constant area (Eq.~(\ref{eq:Helfrich}))~\cite{kraus1995gravity}. Colloidal membranes are porous to the solvent, but with surface-to-surface virus spacing of $\sim$ 10 nm they are impermeable to the depleting polymer (500 kDa dextran, $R_g \approx$~20~nm)~\cite{Balchunas2020}. Therefore, colloidosomes can support differences in osmotic pressure; they can inflate or deflate, which is accounted for by a second Lagrange multiplier enforcing a fixed volume constraint. The colloidosome shape is controlled by two parameters: a dimensionless surface area $\alpha =A (\sigma g / \kappa)^{2/3}$, and an inflation parameter $\lambda = V / V_{0}$. Here, $\sigma$ is the membrane areal mass density, $g$ is gravitational acceleration, $A$ and $V$ are the measured area and volume and $V_{0}$ is the volume of the numerically-computed energy-minimizing contour at zero osmotic pressure difference between interior and exterior. For small $\alpha$, or equivalently small vesicle area, curvature energy dominates leading to sphere-like colloidosomes. For large $\alpha$, the gravitational energy causes sagging deformations (Fig. \ref{fig:Fig1}f). Similarly, deflated colloidosomes ($\lambda<1$) assume more biconcave shapes while inflated ones ($\lambda>1$) are more spherical (Fig. \ref{fig:Fig1}g). 

To test our prediction of colloidsome shapes we first interdependently measured mean curvature $\kappa=11,000 \pm 1000\,k_\text{B} T$, from thermal flucutation  (SI Sec.~1B). Theoretical arguments estimate $\Bar{\kappa} = 50\,k_\text{B} T$ (SI Sec.~1D)~\cite{Khanra2022}. We then imaged colloidosomes using confocal microscopy, extracted their 2D mesh, and measured their shape, area, and volume (Materials and Methods). Our model quantitatively described measured cross-sections without adjustable parameters (Fig. \ref{fig:Fig1}h). The goodness-of-fit average value was $\langle d^2 \rangle / A = 0.015$ and the largest value being $\mathrm{max}\,(\langle d^2 \rangle / A )= 0.05$, where $d$ is the distance between the experimentally measured vesicle contour and the predicted contour (Fig. S5). All colloidosomes were underinflated, which increased with increasing size. We conclude that colloidosomes minimize the Helfrich energy, given a fixed volume and area.

\subsection{Gravity-assisted colloidosome assembly}

\begin{figure*}[t]
\centering
\includegraphics[width=0.85\textwidth]{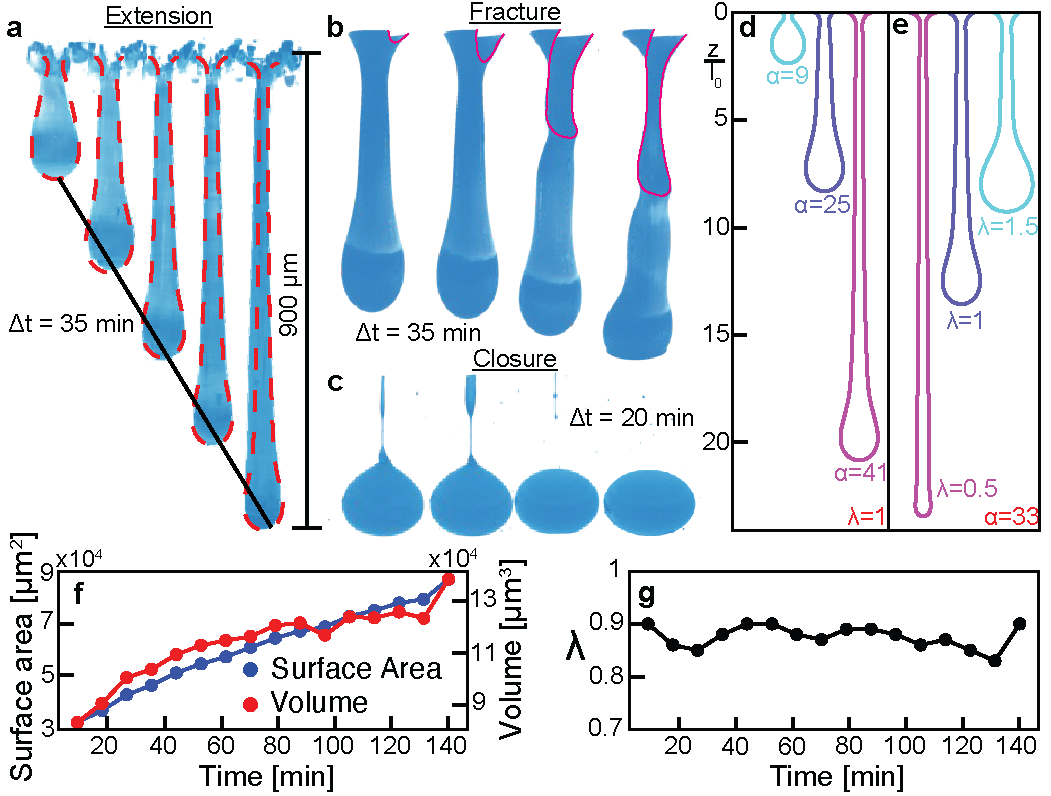}
\caption{
\textbf{Gravity-induced extension of pendent colloidosomes.} \textbf{a,} A pendent colloidosome extends under gravity. Red lines are theory predictions using measured surface area and $\lambda$ as a fitting parameter. \textbf{b,} A pendent colloidosome undergoes fracture. \textbf{c,}  A twisted ribbon-like tether ruptures, leaving behind a closed colloidosome. \textbf{d-e,} Predicted pendent colloidosome shapes with varying $\alpha$ and fixed volume and varying $\lambda$ and fixed $\alpha$ = 33. \textbf{f-g,} Time dependence of  surface area, volume and inflation parameter $\lambda$ for an extending colloidosome. }
\label{fig:Fig2}
\end{figure*}

The slow dynamics and large length scales revealed a multistep kinetic pathway leading to colloidosome formation. Immediately after inverting the chamber, the pendent colloidosomes underwent gravity-induced elongation, producing hollow tube-like tethers (Fig. \ref{fig:Fig2}a), similar to uniform tethers produced by applying a point force to lipid vesicles~\cite{powers2002fluid,roux2002minimal}. The diameter of such tethers decreased with increasing size, due to the gravitational stress increasing up the tube length. The pendent colloidosomes continued extending until either reaching an equilibrium length (Fig. S6) or the chamber bottom. The latter relieved the gravitational stresses and halted extension. A second phase more rapid phase began with the formation of a crack at the membrane's attachment point with the ceiling, where the gravitational stresses were largest (Fig. \ref{fig:Fig2}b). Once nucleated, the crack propagated downward, unwrapping the tube-like tether, leaving a twisted ribbon connecting the partially closed colloidosome to the ceiling (Fig. \ref{fig:Fig2}c)~\cite{Balchunas2020}. This ribbon kept a single pore open. In the final closure phase, the ribbon thinned and twisted over minutes to hours, eventually rupturing. Subsequently, the pore rapidly closed, completing the topological change to a closed colloidosome (Fig. \ref{fig:Fig2}c, Movie S2). 

To understand the dynamics of pendent colloidosome extension, we numerically solved the Helfrich equation with boundary conditions that account for the ceiling attachment (SI Eqs.26-32), and varied dimensionless parameters, $\alpha$ and $\lambda$ (SI Sec.~3B). Large $\alpha$, associated with increasing surface area, produced elongated contours (Fig. \ref{fig:Fig2}d). Similarly, deflated contours extended along the $z$ direction ($\lambda<1$), while inflated contours retracted and became bulbous ($\lambda>1$) (Fig. \ref{fig:Fig2}e). We fitted a time series of extending pendent colloidosomes shapes to theoretical predictions. We measured their areas and volumes, calculating $\alpha$ and fitting contours using $\lambda$ as an adjustable parameter (Fig. \ref{fig:Fig2}f). Predicted contours matched the measurements, indicating that the extending dynamics was sufficiently slow for an extending colloidosome to minimize its energy given the particular area and volume (Fig. \ref{fig:Fig2}a). The area and volume increased throughout the extension (Fig. \ref{fig:Fig2}f), showing that pendent colloidosomes recruited membrane from the attachment at the ceiling. Increasing the volume with a constant mass of enveloped dextran would decrease $\lambda$. However, measured $\lambda$ remained close to unity (Fig. \ref{fig:Fig2}g), suggesting that deflated configurations incur a significant energetic cost, and therefore resist gravity-driven extension. Consequently, the extension rate is determined by the rate at which the dextran can flow through the opening at the ceiling to equilibrate the osmotic pressure difference across the membrane and keep $\lambda$ close to unity.

\subsection{Pathways of colloidosome disassembly}

\begin{figure*}[t]
\centering
\includegraphics[width=0.85\textwidth]{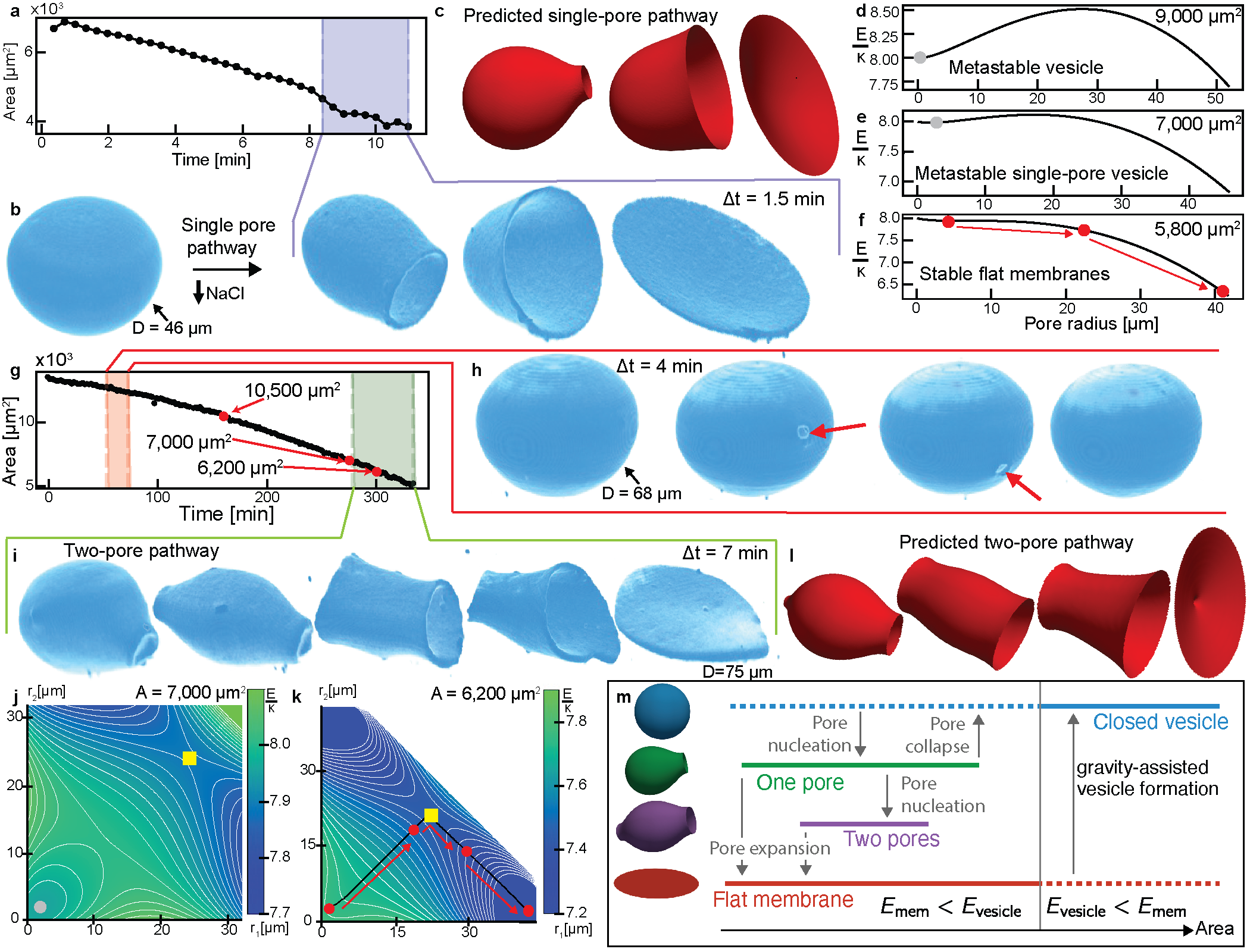}
\caption{\textbf{Vesicle to disk transformation.} \textbf{a,} Decreasing vesicle area after buffer exchange. \textbf{b,} Vesicle unwraps by a single pore pathway. \textbf{c,} Predicted vesicle shapes along one pore pathway. \textbf{d-f,} Single pore energy landscape with decreasing surface areas. Gray points denote local minima. Conformations associated with the red point are shown in c \textbf{g,} Colloidosome area undergoing transient pore opening (red) and the two-pore unwrapping (green). \textbf{h,} Transient pore opening.  \textbf{i,} Vesicle unwrapping via the two-pore pathway.  \textbf{j,k,} Two pore energy landscape for decreasing surface area. The reduced energy $\frac{E}{\kappa}$ as a function of the radii, $r_1, r_2$, of both pores. Gray point denotes local minima; yellow squares denote saddle points. White space in (k) denotes a cutoff in the algorithm at large $r_1$ and $r_2$, where the energy landscape does not influence the shape dynamics. \textbf{l,} Predicted two-pore pathway, following the gradient of steepest descent on the energy landscape in (k). Conformations corresponding to four points along the energy landscape, indicated with red points in (k).  \textbf{m,} Summary of observed membrane states and the transitions between them as a function of area.}
	\label{fig:Fig3}
\end{figure*}

To control colloidosome morphology and topology \emph{in situ} we explored a regime where disks and edgeless vesicles have comparable energetic costs. A buffer exchange device, reduced the salt concentration, increasing electrostatic repulsion between charged rods, causing their evaporation into the background, and decreasing the colloidosome area (Fig. \ref{fig:Fig3}a). Decreasing area increased the concentration and osmotic pressure of interior dextran, producing a pressure difference across the membrane, and increasing its tension. Above a critical value, the fluid membrane ruptured, generating a pore and relieving the osmotic pressure difference. The pore remained open until the decreasing membrane area approached the critical area $A_1^*$. At that point, the pore size increased until the colloidosome transformed into a flat sheet (Fig. \ref{fig:Fig3}b, Movie S3, Movie S4). 

Slowing the rate of area decrease revealed a distinct multistage disassembly pathway (Fig. \ref{fig:Fig3}g). Initially, similar to the first pathway pores nucleated in slowly shrinking vesicles (Fig. \ref{fig:Fig3}h, Movie S5). The concentration difference caused dextran outflow through the pore, reducing the membrane tension, and resealing the pore. As the vesicle continuously shrank, this cycle repeated itself. Similar cascades have been observed in lipid vesicles~\cite{karatekin2003cascades,oglkecka2014oscillatory}. The seconds-long transient pores were analogous to the milliseconds-long dynamics observed in lipid vesicles under osmotic shock, mechanical stress, or electroporation~\cite{Sandre1999,Chabanon2017,Moroz1997,dimova2009vesicles,hamada2010membrane,malik2022pore}. The second stage was initiated as the vesicle approached the critical area, $A_1^*$. In this limit, the pore did not reseal. Rather, a second pore, diametrically opposed to the first, nucleated, generating a topologically distinct intermediate structure with two boundaries. With continued area decrease, both pores grew yielding a cylinder-like intermediate structure. Eventually, the symmetry of the intermediate hollow cylinder-like was broken as one pore grew in size while the other shrank. The smaller pore resealed completing the topological transition into a flat disk (Fig. \ref{fig:Fig3}i, Movie S6). 

To gain insight into disassembly pathways we imaged colloidosome's intermediate shapes. Using the pore openings as boundary conditions, and minimizing the elastic energy predicted the experimentally measured shapes (Fig. S7). Therefore, disassembly follows quasi-static (adiabatic) dynamics that can be understood by the energy landscape.  Motivated by this observation we first calculated the energy associated with a single-opening pathway. For large membranes ($\approx$ 9,000 µm$^2$) the global minimum is at zero pore size, indicating the stability of closed colloidosomes (Fig. \ref{fig:Fig3}d). Decreasing the area to $\approx$ 7,000 µm$^2$ creates a local energy minimum away from zero, indicating a metastable vesicle with one opening (Fig. \ref{fig:Fig3}e). At this point, the global minimum is a flat disk, but the vesicle cannot unwrap due to the energy barrier~\cite{Boal1992}. Decreasing area further ($\leq$ 5,800~$ \text{µm}^2$) eliminates the barrier, at which point the vesicle rapidly unwrapped into a flat membrane (Fig. \ref{fig:Fig3}f). Equally spaced configurations along the energy landscape predicted shapes that agree with the experiment (Fig. \ref{fig:Fig3}c, Movie S4). Experimentally observed rapid increase in pore size occurs at a critical area where the energy barrier disappears ($\approx$ 5,800 µm$^2$) (Fig. S8). 

To understand the two-pore disassembly pathway, we constructed a 2D energy landscape with fixed area $A$, but varying radii $r_1$ and $r_2$ of the two pores (SI Sec.~3E). Large area vesicles ($\approx$ 10,000 µm$^2$) have a global energy minimum at zero pore size, indicating stable colloidosomes. For intermediate areas ($\approx$ 7,000 µm$^2$) the local energy minimum moves away from the origin producing a metastable state with two equal-sized pores (Fig. \ref{fig:Fig3}j). Notably, one-pore configurations are energetically less favorable than the two-pore state explaining the preference for two pores given enough time to nucleate them~\cite{Umeda2005}. Even though the global minimum for these parameters is a flat disk, vesicle unwrapping is suppressed by an energy barrier. With continuously decreasing area ($\approx$ 6,200 µm$^2$) the energy barrier first disappears along the pathway where both $r_1$ and $r_2$ increase concurrently (Fig. \ref{fig:Fig3}k). Once the barrier disappears the steepest gradient descent drives the system toward the saddle point, which corresponds to a tube-like configuration. Subsequently, a spontaneous symmetry breaking expanded one pore while the other one shrank and resealed, thus completing the unwrapping dynamics. Taking equally spaced configurations along a gradient descent of the energy landscape explains experimental observations (Fig. \ref{fig:Fig3}l, Movie S6). Our work shows that vesicle unwrapping is controlled by two time scales; the rate of area change and the time to nucleate a pore. Importantly, the two-pore configuration has lower energy than the single-pore configuration (Fig. S9). Therefore, slowly shrinking vesicles can overcome the energy barrier associated with second-pore nucleation, which lowers the barrier for disassembly.  

\subsection{Size-selective encapsulating colloidosomes} 

\begin{figure}[h]
\centering
\includegraphics[width=0.4\textwidth]{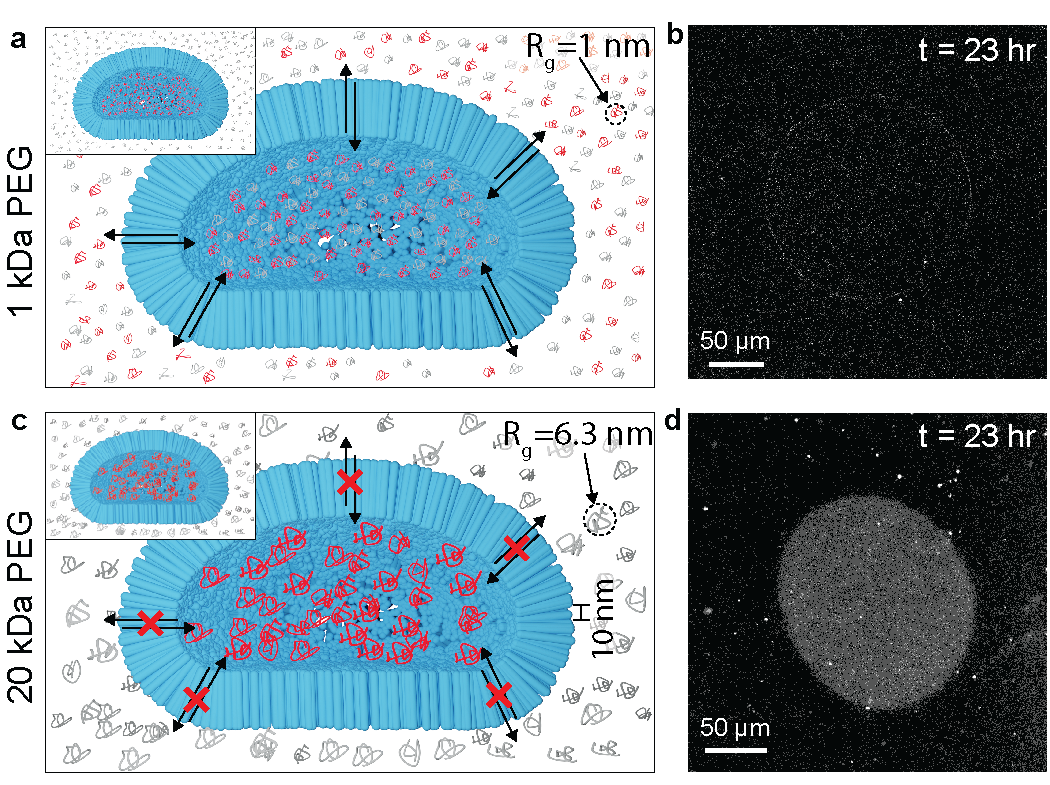}
\caption{\textbf{Encapsulation by colloidosomes.} \textbf{a,} Vesicle containing 1 kDa photoactivatable PEG. Initially (inset), the photoactivated PEG is inside the vesicle. Over time the PEG diffuses through the membrane, mixing the active and inactive PEG. \textbf{b,} Lack of contrast between the interior and exterior indicates that the labeled PEG has diffused after 23 hours. \textbf{c,} Vesicle containing 20 kDa photoactivatable PEG. Initially (inset) the photoactivated PEG (red) is inside the vesicle. The 20 kDa PEG cannot pass through the membrane, so the photoactivated PEG remains separated from the inactive PEG. \textbf{d,} Fluorescence imaging shows that photoactivated PEG is encapsulated in the vesicle for 23 hours.}
\label{fig:Fig4}
\end{figure}

We investigated the colloidosome's ability to selectively encapsulate and release nano-sized cargoes. We included a low-volume fraction of poly(ethylene glycol) (PEG) labeled with a photoactivatable fluorescent tag into the virus-dextran mixture (Materials and Methods). Once vesicles formed, the PEG was photoactivated by UV irradiation, and the fluorescence intensity was visualized. We first used a 1 kDa PEG with a radius-of-gyration $R_g$ $\approx 1$ nm~\cite{Devanand1991}. Twenty-three hours after activation the interior fluorescent intensity of the colloidosome was comparable to the exterior intensity, indicating that the molecules had diffused away through the colloidosome (Figs. \ref{fig:Fig4}a,b). Next, we used 20 kDa PEG with $R_g = 6.3$ nm.  In contrast to the smaller PEG chains, uniform fluorescent signal remained through the colloidosome interior, demonstrating encapsulation (Figs. \ref{fig:Fig4}c,d). The observed size-selective permeability is consistent with previously measured surface-to-surface spacing between the virus rods of $\sim$10 nm~\cite{Balchunas2019}.  These observations demonstrate that colloidosomes are contiguous size-selective structures that can persistently envelop cargoes and lack defects. While previous studies have made nano-porous materials from sheets of hollow viral rods, our colloidosomes are porous due to the interstitial spacing between viral particles~\cite{Zhang2018}. This inter-particle spacing is determined by the balance of electrostatic repulsion and depletion attraction, opening the possibility of using colloidosomes as stimuli-responsive size-selective encapsulating agents.   

\section{Conclusions} 
We elucidated a pathway that robustly transforms 2D disk-shaped colloidal membranes into closed 3D colloidal vesicles or colloidosomes. The gravity-induced shapes of colloidal membranes are quantitatively described by the Helfrich Hamiltonian, a theoretical model that also predicts the behavior of lipid bilayers. Thus, colloidal vesicles reveal universal behavior relevant to all fluid-membrane systems. By balancing edge and bending energy, we prepared marginally stable colloidosomes close to the topological transition associated with their creation and destruction. Experiments and theory revealed that the pathway for vesicle disassembly with the lowest energy barrier involves a transient conformation that is topologically distinct from both the initial and the final states. In comparison, the gravity-induced assembly pathways of colloidosomes are distinct from those found in lipid bilayers. When coupled with methods for controlling colloidal membrane's edge structure and tension, their in-plane phase separation and their Gaussian curvature modulus~\cite{gibaud2012reconfigurable,sharma2014hierarchical,gibaud2017achiral,Khanra2022} advances described here provide a robust platform for forming structured 2D fluid films and controlling their three-dimensional shapes and topologies. Besides fundamental interest, the principles by which the area and edge tension control the vesicle shape and topology, provides a platform for encapsulating, transporting, and delivering nanosized cargoes.

\section{Materials and Methods}
M13KO7 and M13-wt virus were grown using host E. coli strain ER2738, following standard biological protocols~\cite{Wood1983}. Gel electrophoresis revealed that the purified M13-wt virus had a significant amount of end-to-end multimers, which prevents defect-free membrane formation. The multimers were removed using isotropic-nematic phase separation~\cite{Barry2010}. All viruses were suspended in 100 mM NaCl, 20mM Tris-HCl (pH = 8.0) media.

Viruses were labeled with either DyLight 550 or DyLight 488 (Thermo Fisher Scientific) amine reactive dye for the purpose of fluorescence imaging. There are $\sim$3600 and $\sim$2700 labeling sites on M13KO7 and M13-wt rods, respectively. 10\% of the sites were labeled for experiments that image the motion and orientation of individual rods. 1\% of the sites were labeled for all other experiments. 

The number density of viruses in a given suspension was measured with UV-Vis spectrophotometer (Multiscan GO, Thermo Fisher Scientific). The two kinds of viruses were mixed at the desired stoichiometric ratio, and Dextran (MW 500 kDa; Sigma-Aldrich) was added. Coverslips were coated with polyacrylamide brush before sample preparation to prevent membranes from adhering to coverslips. Sample chambers were made from coated coverslips and cleaned slides, using parafilm as spacer. The suspension was injected into the chamber and the chamber was sealed with optical glue (Norland).

Samples were observed using an inverted widefield microscope (Olympus IX83) equipped with 100X oil immersion phase and DIC objectives (UPLanFLN-100X/1.30 Oil Ph3, UPLanFLN-100X/1.30 Oil), motorized z-drive, CCD and EMCCD cameras (Photometrics CoolSNAP HQ2, Andor iXon Ultra 888). A Peltier stage (PE120, Linkam) was used to vary sample temperature. Z-stacks were captured using this microscope in fluorescence mode, followed by deconvolution to represent the structures in 3D qualitatively. Confocal microscope (Zeiss LSM 880 Airyscan, equipped with Plan Apo 63X 1.4NA oil objective) was used for capturing z-stacks for quantitative analysis.

\subsection{Phage production}\label{appendix:makeVesicles}

To produce viral particles, the nano385 phagemid was transfected into cells and then grown in conjunction with a helper phage. The helper phage M13KO7 was grown in E. coli strain ER2738 following previously established protocols~\cite{Wood1983}. After viral proliferation, bacteria were removed from the growth media by centrifugation. The helper phage was precipitated out of the media by adding 20 g/L PEG (8 kDa, Sigma-Aldrich) and 20 g/L NaCl, and pelleted by centrifugation (45,000 xg for 15 mins). The resulting pellet was rinsed and resuspended in virus buffer (2.4 mg/mL Tris-HCl, pH 8.0) at 60 mg/mL to form the helper phage stock solution.

The nano385 plasmid was transformed into a competent F' E. coli strain (NEB C2992H) and grown overnight in 5mL 2XYT starter cultures containing 100 µg/mL ampicillin~\cite{evans1995litmus, Wood1983}. Starter flasks containing 50 mL 2XYT were infected with 1 mL of overnight starter culture and helper phage stock solution was added to a final concentration of 2 µg/mL M13KO7 and incubated for 1 hour. The 1L growth flasks containing 2XYT, 100 µg/mL ampicillin, 25 µg/mL kanamycin, 200 µg/mL MgCl$_2 \cdot$H$_2$O  and 120 µg/mL MgSO$_4$ were infected with 5 mL of the starter flask content, and grown with constant shaking at $37^\circ C$. Both nano385 and M13KO7 are extruded from infected bacteria throughout growth. Upon reaching OD 1.5, flasks were removed from incubation and cooled on ice. 

To purify the nano385 phagemid, bacteria were removed with one round of low-speed centrifugation (10 min at 4,000xg, Fiberlite F9-6 x 1000 LEX fixed angle rotor, Thermo Scientific) followed by a second high-speed round (15 min, 12,000 xg). The supernatant was filtered through a 0.22 µm filter to remove any remaining bacteria. The phage was precipitated by adding 50 g/L PEG 8 kDa and 30 g/L NaCl and pelleted by centrifugation (30 min at 12,000 xg) and resuspended in Tris buffer (2.4 mg/mL tris-HCl, pH 8). To increase the phage purity, we performed two additional centrifugation steps (45,000 xg for 15 minutes) to remove bacterial debris, followed by the addition of 50 g/L PEG 8 kDa and 30 g/L NaCl and a second spin to pellet the phage.

After purification, the virus suspension contained a mixture of both M13K07 and nano385. Adding Dextran (MW $\approx$ 500k, Sigma-Aldrich) to the mixed growth product induces the isotropic to nematic phase transition in M13KO7 at a lower dextran concentration than the nano385. Dextran was added in steps of 5 mg/mL and centrifuged (22,000 xg, 15 minutes), to condense the nematic phase at the bottom of the centrifuge tube at each step. Gel electrophoresis was used to confirm the separation of M13KO7 and nano385 for each fraction. The fractions that contained only nano385 (typically the 45 mg/mL fraction) were centrifuged (280,000 xg, 1 hr) to pellet the phage and resuspended in nano385 buffer (7.3 mg/mL NaCl, 2.4 mg/mL Tris-HCl, pH 8.0). Optionally, an additional purification step using anion exchange chromatography could be performed (POROS, GoPure XQ)~\cite{Monjezi2010}. This procedure dramatically increased membrane size and the frequency of colloidosome formation [Movie S1].

\subsection{Phage labeling}\label{appendix:LabelingPhages}

To label phages for fluorescent microscopy, we labeled the primary amines of the virus major coat protein with an amine-reactive fluorophore (DyLight-NHS ester 550; Thermo Fisher) according to the dye manufacturer's instructions. Each virus was labeled at a low percent fraction ($\approx 5 \%$ of $\approx 1,150$ major coat proteins per phage) to ensure the fluorescent dye does not alter the membrane properties. The labeling percentage was confirmed by spectrophotometer (Nanodrop One; Thermo Fisher).

\subsection{Sample preparation}

Samples comprised fluorescently labeled nano385 and 500 kDa Dextran in nano385 buffer (7.3 mg/mL NaCl, 2.4 mg/mL tris-HCl, pH 8).  Dextran acts as a polymer depletant, inducing an entropic attraction between phages~\cite{Barry2010}. Low-polydispersity Dextran (MW~500,000 Da, Sigma-Aldrich) was made using ethanol precipitation which produced significantly cleaner and larger membranes, resulting in more frequent colloidosome formation. To fractionate, ethanol was added dropwise to a solution of 0.2\% Dextran under vigorous stirring at 23 $^\circ$C. After reaching 31\% (w/w) ethanol, Dextran precipitates were removed by centrifugation (20 min at 17,000g, Fiberlite F9-6 x 1000 LEX fixed angle rotor, Thermo Scientific). Ethanol was then added to a concentration of 32\% (w/w) and the precipitate was again collected by centrifugation. The solvent was removed by freeze-dry lyophilization, and the resulting powdered Dextran was reconstituted in nano385 buffer to a concentration of 100 mg/mL Dextran.

Chambers were made using a microscope slide and coverslip that were coated with acrylamide brush to suppress adsorption and separated by parafilm spacers~\cite{Lau2009}. The number of parafilm layers set the chamber thickness. Four parafilm layers generated $\approx 500$ µm thick chambers which consistently yielded closed colloidosomes. To study the extension dynamics of pendent colloidosome 3 mm thick chambers were used. Due to membrane sedimentation, 3 mm chambers required a lower final phagemid concentration (0.4 mg/mL) than 500 $\mu m$ chambers (1 mg/mL) to achieve a similar density of membranes at the surface. All samples were prepared with the final concentration of 54 mg/mL fractionated Dextran. 

After preparation, the sample was stored overnight with the coverslip facing upward. The next day, curved membranes layered the microscope-glass-side of the chamber (Fig. S4). The sample was then inverted so that the coverslip faced downward. To image membrane extension and tearing, samples were imaged immediately. To image fully formed colloidosomes, the chamber was left inverted for a day or more, and the resulting colloidosomes were imaged. 

\subsection{Production of vesicles from shorter phagemids}

To produce vesicles from viral particles shorter than 385 nm, we decreased the length of the phagmid DNA. To produce viral particles that were $\approx$315 nm in length, we removed the LacZ$\alpha$ site from the nano385 plasmid, shortening the length of the phagemid from 2814 bp to 2302. This 19\% reduction in phagemid length shortened the length of the resulting phage by 19\%, resulting in  phage of estimated length 315 nm. The phagemid was then transformed into a competent E. Coli strain, and grown and purified according to an identical protocol as nano385. When making samples, a higher dextran concentration of 65 mg/mL was required to form membranes.

To make phages that were 200 nm in length, we used the pScaf-1512-1 phagemid [Addgene \#111402] along with the helper plasmid HP17K07 [Addgene \#120346] in the bacterial host XL1Blue~\cite{pScaf2017, pScaf2018}. These were grown as described previously. Vesicles were formed at 77.1 mg/mL dextran and in a buffer of 7.89 mg/mL NaCl, 2.4 mg/mL Tris-HCl, pH 8.0. 

\subsection{Dialysis chamber}

To perform buffer exchange experiments, we built a custom dialysis device suitable for vesicle formation. Two holes were drilled into a standard microscope glass slide, which was then coated with an acrylamide brush~\cite{Lau2009}. A 20 kDa dialysis membrane was glued on the top surface of the slide, covering the drilled holes. A PDMS buffer exchange chamber was then glued over the top of the dialysis membrane. On the opposing side of the microscope slide, a chamber was formed using a coverslip with 500 µm of parafilm as a spacer. The bottom chamber was filled with the sample (0.4 mg/mL nano385, 54 mg/mL dextran in 7.3 mg/mL NaCl, 20 2.4 mg/mL, pH 8), and sealed with Norland optical adhesive [NOA 61] glue. The top buffer-exchange chamber was filled with nano385 buffer (7.3 mg/mL NaCl, 2.4 mg/mL, pH 8), and sealed with a flexible epoxy. The sample was stored overnight with the coverslip facing up to form membranes on the microscope glass. The chamber was then inverted so that the coverslip faced downward and was stored for an additional day, to form colloidosomes on the bottom coverslip. The flexible epoxy covering the holes in the dialysis chamber was removed, and the chamber was flushed and filled with 2.4 mg/mL Tris-HCl, pH 8 buffer. Salt diffused out of the sample chamber over the course of several hours, while the colloidosome shape change was imaged from below. 

\subsection{Imaging colloidosomes}\label{appendix:ImagingVesicles}

To image the fluorescent membranes, we used a spinning disk confocal microscope (Crest X-Light V2) and a Hamamatsu ORCA-Flash4.0 V3 attached to a Nikon Ti2 base. To achieve the speed of imaging required for 3D scans, the camera and the spinning disk were triggered using a Nikon Breakout Box (NI-BB). All images were taken using a water immersion objective to minimize axial distortion. To image sedimented colloidosomes or colloidosome disassembly we used a short-working distance, high NA objective (N40XLWD-NIR, Nikon). To image the several millimeter depth required to study colloidosome extension, we mounted a long-range objective scanner on the scope (V-308 Voice Coil, Physik Instrumente) using a custom bracket, along with a long-working distance water dipping objective (N40X-NIR, Nikon).

\subsection{Encapsulation experiments}

For encapsulation experiments, we prepared vesicle samples as described previously, with the addition of 5 µg/mL photoactivatable PEG. Photoactivatable PEG was made in lab by conjugating amine-PEG of desired molecular weight with PA Janelia Fluor 646 NHS Ester. Unconjugated dye was removed by overnight dialysis against nano385 buffer, and flash frozen in a stock concentration of 50 µg/mL. Vesicles were formed as before. The PA-PEG was activated using the 365 nm laser line on a Crest X-Light V2 spinning disk microscope, with a 40x objective. The sample was then monitored for fluorescence over the next day. 

\subsection{Contouring colloidosomes}\label{appendix:Contouring}

To contour colloidosomes, the 3D confocal images were imported into Fiji~\cite{Fiji2012}. Initial rough contours of fluorescent objects in each image were found using the Ridge Detector plugin~\cite{Steger1998}. This generated a mask that roughly contoured the colloidosome, but included noise and other membrane objects. To filter these out, we recognize that the z-scans of the colloidosomes are composed of roughly-circular cross sections. We then fit each ridge detection image in the z-scan with a circle fit using a RANSAC algorithm in Python. This picked out colloidosome-like objects while rejecting noise and line-like cross-sections that make up other membranes. The Z-stacks of the circular fits acted as an initial point cloud which was converted into a mesh using MeshLab~\cite{MeshLab}. This point cloud was cleaned from outliers using the filters "simplify point cloud" and "Compute Normal for Point Set". These points were then used to construct a surface using the filter "Ball Pivoting". Each mesh was then visually inspected, and self-intersections or holes were manually repaired. 

The cleaned meshes were then put through a final round of processing. In the above-described contouring scheme, colloidosomes were assumed to have perfectly circular cross-sections, which introduces unrealistic constraints on the final contour and leads to artifacts. To remedy this, we developed an iterative algorithm to evolve the initial mesh by attracting it toward regions of high intensity. To begin, first-order directional derivatives of the image were taken using a difference-of-Gaussian filter. Each point on the mesh is acted on by a force, $\overrightarrow{F} = \xi \overrightarrow{\nabla} I$, where $\xi$ is a tuned constant and $I$ is the image intensity. This acts to draw each point on the mesh surface toward local regions of high intensity. This algorithm was iterated until convergence. In practice, with the proper choice of $\xi$, each point moved only several pixels, settling within one hundred iterations. The area and volume of the colloidosomes were then measured from the mesh, using the filter "Compute Geometric Measures".

\bigskip

\begin{acknowledgments}

This research was primarily supported by the National Science Foundation through a grant from NSF-BMAT. P.S. and A.K. acknowledge funding from DST-SERB grants CRG/2019/000855 and WEA/2023/000006. F.B. acknowledges the support of the GBMF post-doctoral fellowship (under grant \#2919). R.A. acknowledges support by Schmidt Science Fellows, in partnership with the Rhodes Trust. S.S. acknowledges support from the HFSP cross-disciplinary fellowship LT0003/2023-C. R. A. P. and T. R. P. acknowledge support from National Science Foundation Grant CMMI-2020098, and T. R. P. acknowledges support from NSF Materials Research Science and Engineering Centers Grant DMR-2011846. We thank Federico Cao for the valuable discussions. Certain equipment, instruments, software, or materials are identified in this paper in order to specify the experimental procedure adequately.  Such identification is not intended to imply recommendation or endorsement of any product or service by NIST, nor is it intended to imply that the materials or equipment identified are necessarily the best available for the purpose.
\end{acknowledgments}

\bibliography{refs}

\begin{thebibliography}{70}%
\makeatletter
\providecommand \@ifxundefined [1]{%
 \@ifx{#1\undefined}
}%
\providecommand \@ifnum [1]{%
 \ifnum #1\expandafter \@firstoftwo
 \else \expandafter \@secondoftwo
 \fi
}%
\providecommand \@ifx [1]{%
 \ifx #1\expandafter \@firstoftwo
 \else \expandafter \@secondoftwo
 \fi
}%
\providecommand \natexlab [1]{#1}%
\providecommand \enquote  [1]{``#1''}%
\providecommand \bibnamefont  [1]{#1}%
\providecommand \bibfnamefont [1]{#1}%
\providecommand \citenamefont [1]{#1}%
\providecommand \href@noop [0]{\@secondoftwo}%
\providecommand \href [0]{\begingroup \@sanitize@url \@href}%
\providecommand \@href[1]{\@@startlink{#1}\@@href}%
\providecommand \@@href[1]{\endgroup#1\@@endlink}%
\providecommand \@sanitize@url [0]{\catcode `\\12\catcode `\$12\catcode `\&12\catcode `\#12\catcode `\^12\catcode `\_12\catcode `\%12\relax}%
\providecommand \@@startlink[1]{}%
\providecommand \@@endlink[0]{}%
\providecommand \url  [0]{\begingroup\@sanitize@url \@url }%
\providecommand \@url [1]{\endgroup\@href {#1}{\urlprefix }}%
\providecommand \urlprefix  [0]{URL }%
\providecommand \Eprint [0]{\href }%
\providecommand \doibase [0]{https://doi.org/}%
\providecommand \selectlanguage [0]{\@gobble}%
\providecommand \bibinfo  [0]{\@secondoftwo}%
\providecommand \bibfield  [0]{\@secondoftwo}%
\providecommand \translation [1]{[#1]}%
\providecommand \BibitemOpen [0]{}%
\providecommand \bibitemStop [0]{}%
\providecommand \bibitemNoStop [0]{.\EOS\space}%
\providecommand \EOS [0]{\spacefactor3000\relax}%
\providecommand \BibitemShut  [1]{\csname bibitem#1\endcsname}%
\let\auto@bib@innerbib\@empty
\bibitem [{\citenamefont {Nath}\ \emph {et~al.}(2003)\citenamefont {Nath}, \citenamefont {Crawford}, \citenamefont {Carpenter},\ and\ \citenamefont {Coen}}]{Nath2003}%
  \BibitemOpen
  \bibfield  {author} {\bibinfo {author} {\bibfnamefont {U.}~\bibnamefont {Nath}}, \bibinfo {author} {\bibfnamefont {B.~C.~W.}\ \bibnamefont {Crawford}}, \bibinfo {author} {\bibfnamefont {R.}~\bibnamefont {Carpenter}},\ and\ \bibinfo {author} {\bibfnamefont {E.}~\bibnamefont {Coen}},\ }\bibfield  {title} {\bibinfo {title} {Genetic control of surface curvature.},\ }\href@noop {} {\bibfield  {journal} {\bibinfo  {journal} {Science}\ }\textbf {\bibinfo {volume} {299}},\ \bibinfo {pages} {1404} (\bibinfo {year} {2003})}\BibitemShut {NoStop}%
\bibitem [{\citenamefont {Liang}\ and\ \citenamefont {Mahadevan}(2011)}]{Liang2011}%
  \BibitemOpen
  \bibfield  {author} {\bibinfo {author} {\bibfnamefont {H.}~\bibnamefont {Liang}}\ and\ \bibinfo {author} {\bibfnamefont {L.}~\bibnamefont {Mahadevan}},\ }\bibfield  {title} {\bibinfo {title} {Growth, geometry, and mechanics of a blooming lily.},\ }\href@noop {} {\bibfield  {journal} {\bibinfo  {journal} {Proceedings of the National Academy of Sciences}\ }\textbf {\bibinfo {volume} {108}},\ \bibinfo {pages} {5516} (\bibinfo {year} {2011})}\BibitemShut {NoStop}%
\bibitem [{\citenamefont {Metzger}\ \emph {et~al.}(2008)\citenamefont {Metzger}, \citenamefont {Klein}, \citenamefont {Martin},\ and\ \citenamefont {Krasnow}}]{Metzger2008}%
  \BibitemOpen
  \bibfield  {author} {\bibinfo {author} {\bibfnamefont {R.~J.}\ \bibnamefont {Metzger}}, \bibinfo {author} {\bibfnamefont {O.~D.}\ \bibnamefont {Klein}}, \bibinfo {author} {\bibfnamefont {G.~R.}\ \bibnamefont {Martin}},\ and\ \bibinfo {author} {\bibfnamefont {M.~A.}\ \bibnamefont {Krasnow}},\ }\bibfield  {title} {\bibinfo {title} {The branching programme of mouse lung development.},\ }\href@noop {} {\bibfield  {journal} {\bibinfo  {journal} {Nature}\ }\textbf {\bibinfo {volume} {453}},\ \bibinfo {pages} {745} (\bibinfo {year} {2008})}\BibitemShut {NoStop}%
\bibitem [{\citenamefont {Mitchell}\ \emph {et~al.}(2022)\citenamefont {Mitchell}, \citenamefont {Cislo}, \citenamefont {Shankar}, \citenamefont {Lin}, \citenamefont {Shraiman},\ and\ \citenamefont {Streichan}}]{Mitchell2022}%
  \BibitemOpen
  \bibfield  {author} {\bibinfo {author} {\bibfnamefont {N.~P.}\ \bibnamefont {Mitchell}}, \bibinfo {author} {\bibfnamefont {D.~J.}\ \bibnamefont {Cislo}}, \bibinfo {author} {\bibfnamefont {S.}~\bibnamefont {Shankar}}, \bibinfo {author} {\bibfnamefont {Y.}~\bibnamefont {Lin}}, \bibinfo {author} {\bibfnamefont {B.~I.}\ \bibnamefont {Shraiman}},\ and\ \bibinfo {author} {\bibfnamefont {S.~J.}\ \bibnamefont {Streichan}},\ }\bibfield  {title} {\bibinfo {title} {Visceral organ morphogenesis via calcium-patterned muscle constrictions.},\ }\href@noop {} {\bibfield  {journal} {\bibinfo  {journal} {eLife}\ }\textbf {\bibinfo {volume} {11}} (\bibinfo {year} {2022})}\BibitemShut {NoStop}%
\bibitem [{\citenamefont {Kozlovsky}\ and\ \citenamefont {Kozlov}(2002)}]{kozlovsky2002stalk}%
  \BibitemOpen
  \bibfield  {author} {\bibinfo {author} {\bibfnamefont {Y.}~\bibnamefont {Kozlovsky}}\ and\ \bibinfo {author} {\bibfnamefont {M.~M.}\ \bibnamefont {Kozlov}},\ }\bibfield  {title} {\bibinfo {title} {Stalk model of membrane fusion: solution of energy crisis},\ }\href@noop {} {\bibfield  {journal} {\bibinfo  {journal} {Biophysical journal}\ }\textbf {\bibinfo {volume} {82}},\ \bibinfo {pages} {882} (\bibinfo {year} {2002})}\BibitemShut {NoStop}%
\bibitem [{\citenamefont {Conner}\ and\ \citenamefont {Schmid}(2003)}]{conner2003regulated}%
  \BibitemOpen
  \bibfield  {author} {\bibinfo {author} {\bibfnamefont {S.~D.}\ \bibnamefont {Conner}}\ and\ \bibinfo {author} {\bibfnamefont {S.~L.}\ \bibnamefont {Schmid}},\ }\bibfield  {title} {\bibinfo {title} {Regulated portals of entry into the cell},\ }\href@noop {} {\bibfield  {journal} {\bibinfo  {journal} {Nature}\ }\textbf {\bibinfo {volume} {422}},\ \bibinfo {pages} {37} (\bibinfo {year} {2003})}\BibitemShut {NoStop}%
\bibitem [{\citenamefont {Jahn}\ \emph {et~al.}(2003)\citenamefont {Jahn}, \citenamefont {Lang},\ and\ \citenamefont {S{\"u}dhof}}]{jahn2003membrane}%
  \BibitemOpen
  \bibfield  {author} {\bibinfo {author} {\bibfnamefont {R.}~\bibnamefont {Jahn}}, \bibinfo {author} {\bibfnamefont {T.}~\bibnamefont {Lang}},\ and\ \bibinfo {author} {\bibfnamefont {T.~C.}\ \bibnamefont {S{\"u}dhof}},\ }\bibfield  {title} {\bibinfo {title} {Membrane fusion},\ }\href@noop {} {\bibfield  {journal} {\bibinfo  {journal} {Cell}\ }\textbf {\bibinfo {volume} {112}},\ \bibinfo {pages} {519} (\bibinfo {year} {2003})}\BibitemShut {NoStop}%
\bibitem [{\citenamefont {Nixon-Abell}\ \emph {et~al.}(2016)\citenamefont {Nixon-Abell}, \citenamefont {Obara}, \citenamefont {Weigel}, \citenamefont {Li}, \citenamefont {Legant}, \citenamefont {Xu}, \citenamefont {Pasolli}, \citenamefont {Harvey}, \citenamefont {Hess}, \citenamefont {Betzig} \emph {et~al.}}]{nixon2016increased}%
  \BibitemOpen
  \bibfield  {author} {\bibinfo {author} {\bibfnamefont {J.}~\bibnamefont {Nixon-Abell}}, \bibinfo {author} {\bibfnamefont {C.~J.}\ \bibnamefont {Obara}}, \bibinfo {author} {\bibfnamefont {A.~V.}\ \bibnamefont {Weigel}}, \bibinfo {author} {\bibfnamefont {D.}~\bibnamefont {Li}}, \bibinfo {author} {\bibfnamefont {W.~R.}\ \bibnamefont {Legant}}, \bibinfo {author} {\bibfnamefont {C.~S.}\ \bibnamefont {Xu}}, \bibinfo {author} {\bibfnamefont {H.~A.}\ \bibnamefont {Pasolli}}, \bibinfo {author} {\bibfnamefont {K.}~\bibnamefont {Harvey}}, \bibinfo {author} {\bibfnamefont {H.~F.}\ \bibnamefont {Hess}}, \bibinfo {author} {\bibfnamefont {E.}~\bibnamefont {Betzig}}, \emph {et~al.},\ }\bibfield  {title} {\bibinfo {title} {Increased spatiotemporal resolution reveals highly dynamic dense tubular matrices in the peripheral {ER}},\ }\href@noop {} {\bibfield  {journal} {\bibinfo  {journal} {Science}\ }\textbf {\bibinfo {volume} {354}},\ \bibinfo {pages} {aaf3928} (\bibinfo {year} {2016})}\BibitemShut {NoStop}%
\bibitem [{\citenamefont {Shin}\ \emph {et~al.}(2018)\citenamefont {Shin}, \citenamefont {Ge}, \citenamefont {Arpino}, \citenamefont {Villarreal}, \citenamefont {Hamid}, \citenamefont {Liu}, \citenamefont {Zhao}, \citenamefont {Wen}, \citenamefont {Chiang},\ and\ \citenamefont {Wu}}]{shin2018visualization}%
  \BibitemOpen
  \bibfield  {author} {\bibinfo {author} {\bibfnamefont {W.}~\bibnamefont {Shin}}, \bibinfo {author} {\bibfnamefont {L.}~\bibnamefont {Ge}}, \bibinfo {author} {\bibfnamefont {G.}~\bibnamefont {Arpino}}, \bibinfo {author} {\bibfnamefont {S.~A.}\ \bibnamefont {Villarreal}}, \bibinfo {author} {\bibfnamefont {E.}~\bibnamefont {Hamid}}, \bibinfo {author} {\bibfnamefont {H.}~\bibnamefont {Liu}}, \bibinfo {author} {\bibfnamefont {W.-D.}\ \bibnamefont {Zhao}}, \bibinfo {author} {\bibfnamefont {P.~J.}\ \bibnamefont {Wen}}, \bibinfo {author} {\bibfnamefont {H.-C.}\ \bibnamefont {Chiang}},\ and\ \bibinfo {author} {\bibfnamefont {L.-G.}\ \bibnamefont {Wu}},\ }\bibfield  {title} {\bibinfo {title} {Visualization of membrane pore in live cells reveals a dynamic-pore theory governing fusion and endocytosis},\ }\href@noop {} {\bibfield  {journal} {\bibinfo  {journal} {Cell}\ }\textbf {\bibinfo {volume} {173}},\ \bibinfo {pages} {934} (\bibinfo {year} {2018})}\BibitemShut {NoStop}%
\bibitem [{\citenamefont {Klein}\ \emph {et~al.}(2007)\citenamefont {Klein}, \citenamefont {Efrati},\ and\ \citenamefont {Sharon}}]{Klein2007}%
  \BibitemOpen
  \bibfield  {author} {\bibinfo {author} {\bibfnamefont {Y.}~\bibnamefont {Klein}}, \bibinfo {author} {\bibfnamefont {E.}~\bibnamefont {Efrati}},\ and\ \bibinfo {author} {\bibfnamefont {E.}~\bibnamefont {Sharon}},\ }\bibfield  {title} {\bibinfo {title} {Shaping of elastic sheets by prescription of non-euclidean metrics.},\ }\href@noop {} {\bibfield  {journal} {\bibinfo  {journal} {Science}\ }\textbf {\bibinfo {volume} {315}} (\bibinfo {year} {2007})}\BibitemShut {NoStop}%
\bibitem [{\citenamefont {Armon}\ \emph {et~al.}(2011)\citenamefont {Armon}, \citenamefont {Efrati}, \citenamefont {Kupferman},\ and\ \citenamefont {Sharon}}]{armon2011geometry}%
  \BibitemOpen
  \bibfield  {author} {\bibinfo {author} {\bibfnamefont {S.}~\bibnamefont {Armon}}, \bibinfo {author} {\bibfnamefont {E.}~\bibnamefont {Efrati}}, \bibinfo {author} {\bibfnamefont {R.}~\bibnamefont {Kupferman}},\ and\ \bibinfo {author} {\bibfnamefont {E.}~\bibnamefont {Sharon}},\ }\bibfield  {title} {\bibinfo {title} {Geometry and mechanics in the opening of chiral seed pods},\ }\href@noop {} {\bibfield  {journal} {\bibinfo  {journal} {Science}\ }\textbf {\bibinfo {volume} {333}},\ \bibinfo {pages} {1726} (\bibinfo {year} {2011})}\BibitemShut {NoStop}%
\bibitem [{\citenamefont {Kim}\ \emph {et~al.}(2012)\citenamefont {Kim}, \citenamefont {Hanna}, \citenamefont {Byun}, \citenamefont {Santangelo},\ and\ \citenamefont {Hayward}}]{kim2012designing}%
  \BibitemOpen
  \bibfield  {author} {\bibinfo {author} {\bibfnamefont {J.}~\bibnamefont {Kim}}, \bibinfo {author} {\bibfnamefont {J.~A.}\ \bibnamefont {Hanna}}, \bibinfo {author} {\bibfnamefont {M.}~\bibnamefont {Byun}}, \bibinfo {author} {\bibfnamefont {C.~D.}\ \bibnamefont {Santangelo}},\ and\ \bibinfo {author} {\bibfnamefont {R.~C.}\ \bibnamefont {Hayward}},\ }\bibfield  {title} {\bibinfo {title} {Designing responsive buckled surfaces by halftone gel lithography},\ }\href@noop {} {\bibfield  {journal} {\bibinfo  {journal} {Science}\ }\textbf {\bibinfo {volume} {335}},\ \bibinfo {pages} {1201} (\bibinfo {year} {2012})}\BibitemShut {NoStop}%
\bibitem [{\citenamefont {Sydney~Gladman}\ \emph {et~al.}(2016)\citenamefont {Sydney~Gladman}, \citenamefont {Matsumoto}, \citenamefont {Nuzzo}, \citenamefont {Mahadevan},\ and\ \citenamefont {Lewis}}]{sydney2016biomimetic}%
  \BibitemOpen
  \bibfield  {author} {\bibinfo {author} {\bibfnamefont {A.}~\bibnamefont {Sydney~Gladman}}, \bibinfo {author} {\bibfnamefont {E.~A.}\ \bibnamefont {Matsumoto}}, \bibinfo {author} {\bibfnamefont {R.~G.}\ \bibnamefont {Nuzzo}}, \bibinfo {author} {\bibfnamefont {L.}~\bibnamefont {Mahadevan}},\ and\ \bibinfo {author} {\bibfnamefont {J.~A.}\ \bibnamefont {Lewis}},\ }\bibfield  {title} {\bibinfo {title} {Biomimetic 4d printing},\ }\href@noop {} {\bibfield  {journal} {\bibinfo  {journal} {Nature materials}\ }\textbf {\bibinfo {volume} {15}},\ \bibinfo {pages} {413} (\bibinfo {year} {2016})}\BibitemShut {NoStop}%
\bibitem [{\citenamefont {Evans}\ and\ \citenamefont {Rawicz}(1990)}]{evans1990entropy}%
  \BibitemOpen
  \bibfield  {author} {\bibinfo {author} {\bibfnamefont {E.}~\bibnamefont {Evans}}\ and\ \bibinfo {author} {\bibfnamefont {W.}~\bibnamefont {Rawicz}},\ }\bibfield  {title} {\bibinfo {title} {Entropy-driven tension and bending elasticity in condensed-fluid membranes},\ }\href@noop {} {\bibfield  {journal} {\bibinfo  {journal} {Physical review letters}\ }\textbf {\bibinfo {volume} {64}},\ \bibinfo {pages} {2094} (\bibinfo {year} {1990})}\BibitemShut {NoStop}%
\bibitem [{\citenamefont {Seifert}(1997)}]{seifert1997configurations}%
  \BibitemOpen
  \bibfield  {author} {\bibinfo {author} {\bibfnamefont {U.}~\bibnamefont {Seifert}},\ }\bibfield  {title} {\bibinfo {title} {Configurations of fluid membranes and vesicles},\ }\href@noop {} {\bibfield  {journal} {\bibinfo  {journal} {Advances in physics}\ }\textbf {\bibinfo {volume} {46}},\ \bibinfo {pages} {13} (\bibinfo {year} {1997})}\BibitemShut {NoStop}%
\bibitem [{\citenamefont {Discher}\ \emph {et~al.}(1999)\citenamefont {Discher}, \citenamefont {Won}, \citenamefont {Ege}, \citenamefont {Lee}, \citenamefont {Bates}, \citenamefont {Discher},\ and\ \citenamefont {Hammer}}]{Discher1999}%
  \BibitemOpen
  \bibfield  {author} {\bibinfo {author} {\bibfnamefont {B.~M.}\ \bibnamefont {Discher}}, \bibinfo {author} {\bibfnamefont {Y.-Y.}\ \bibnamefont {Won}}, \bibinfo {author} {\bibfnamefont {D.~S.}\ \bibnamefont {Ege}}, \bibinfo {author} {\bibfnamefont {J.~C.-M.}\ \bibnamefont {Lee}}, \bibinfo {author} {\bibfnamefont {F.~S.}\ \bibnamefont {Bates}}, \bibinfo {author} {\bibfnamefont {D.~E.}\ \bibnamefont {Discher}},\ and\ \bibinfo {author} {\bibfnamefont {D.~A.}\ \bibnamefont {Hammer}},\ }\bibfield  {title} {\bibinfo {title} {Polymersomes: Tough vesicles made from diblock copolymers},\ }\bibfield  {journal} {\bibinfo  {journal} {Science}\ }\textbf {\bibinfo {volume} {284}},\ \href {https://doi.org/10.1126/science.284.5417.1143} {10.1126/science.284.5417.1143} (\bibinfo {year} {1999})\BibitemShut {NoStop}%
\bibitem [{\citenamefont {Dinsmore}\ \emph {et~al.}(2002)\citenamefont {Dinsmore}, \citenamefont {Hsu}, \citenamefont {Nikolaides}, \citenamefont {Marquez}, \citenamefont {Bausch},\ and\ \citenamefont {Weitz}}]{Dinsmore2002}%
  \BibitemOpen
  \bibfield  {author} {\bibinfo {author} {\bibfnamefont {A.~D.}\ \bibnamefont {Dinsmore}}, \bibinfo {author} {\bibfnamefont {M.~F.}\ \bibnamefont {Hsu}}, \bibinfo {author} {\bibfnamefont {M.~G.}\ \bibnamefont {Nikolaides}}, \bibinfo {author} {\bibfnamefont {M.}~\bibnamefont {Marquez}}, \bibinfo {author} {\bibfnamefont {A.~R.}\ \bibnamefont {Bausch}},\ and\ \bibinfo {author} {\bibfnamefont {D.~A.}\ \bibnamefont {Weitz}},\ }\bibfield  {title} {\bibinfo {title} {Colloidosomes: Selectively permeable capsules composed of colloidal particles},\ }\bibfield  {journal} {\bibinfo  {journal} {Science}\ }\textbf {\bibinfo {volume} {298}},\ \href {https://doi.org/10.1126/science.1074868} {10.1126/science.1074868} (\bibinfo {year} {2002})\BibitemShut {NoStop}%
\bibitem [{\citenamefont {Baumgart}\ \emph {et~al.}(2003)\citenamefont {Baumgart}, \citenamefont {Hess},\ and\ \citenamefont {Webb}}]{baumgart2003imaging}%
  \BibitemOpen
  \bibfield  {author} {\bibinfo {author} {\bibfnamefont {T.}~\bibnamefont {Baumgart}}, \bibinfo {author} {\bibfnamefont {S.~T.}\ \bibnamefont {Hess}},\ and\ \bibinfo {author} {\bibfnamefont {W.~W.}\ \bibnamefont {Webb}},\ }\bibfield  {title} {\bibinfo {title} {Imaging coexisting fluid domains in biomembrane models coupling curvature and line tension},\ }\href@noop {} {\bibfield  {journal} {\bibinfo  {journal} {Nature}\ }\textbf {\bibinfo {volume} {425}},\ \bibinfo {pages} {821} (\bibinfo {year} {2003})}\BibitemShut {NoStop}%
\bibitem [{\citenamefont {Xu}\ \emph {et~al.}(2021)\citenamefont {Xu}, \citenamefont {Hueckel}, \citenamefont {Irvine},\ and\ \citenamefont {Sacanna}}]{xu2021transmembrane}%
  \BibitemOpen
  \bibfield  {author} {\bibinfo {author} {\bibfnamefont {Z.}~\bibnamefont {Xu}}, \bibinfo {author} {\bibfnamefont {T.}~\bibnamefont {Hueckel}}, \bibinfo {author} {\bibfnamefont {W.~T.}\ \bibnamefont {Irvine}},\ and\ \bibinfo {author} {\bibfnamefont {S.}~\bibnamefont {Sacanna}},\ }\bibfield  {title} {\bibinfo {title} {Transmembrane transport in inorganic colloidal cell-mimics},\ }\href@noop {} {\bibfield  {journal} {\bibinfo  {journal} {Nature}\ }\textbf {\bibinfo {volume} {597}},\ \bibinfo {pages} {220} (\bibinfo {year} {2021})}\BibitemShut {NoStop}%
\bibitem [{\citenamefont {Umeda}\ \emph {et~al.}(2005)\citenamefont {Umeda}, \citenamefont {Suezaki}, \citenamefont {Takiguchi},\ and\ \citenamefont {Hotani}}]{Umeda2005}%
  \BibitemOpen
  \bibfield  {author} {\bibinfo {author} {\bibfnamefont {T.}~\bibnamefont {Umeda}}, \bibinfo {author} {\bibfnamefont {Y.}~\bibnamefont {Suezaki}}, \bibinfo {author} {\bibfnamefont {K.}~\bibnamefont {Takiguchi}},\ and\ \bibinfo {author} {\bibfnamefont {H.}~\bibnamefont {Hotani}},\ }\bibfield  {title} {\bibinfo {title} {Theoretical analysis of opening-up vesicles with single and two holes.},\ }\href@noop {} {\bibfield  {journal} {\bibinfo  {journal} {Physical Review E—Statistical, Nonlinear, and Soft Matter Physics}\ }\textbf {\bibinfo {volume} {1}} (\bibinfo {year} {2005})}\BibitemShut {NoStop}%
\bibitem [{\citenamefont {Noguchi}\ and\ \citenamefont {Gompper}(2006)}]{noguchi2006dynamics}%
  \BibitemOpen
  \bibfield  {author} {\bibinfo {author} {\bibfnamefont {H.}~\bibnamefont {Noguchi}}\ and\ \bibinfo {author} {\bibfnamefont {G.}~\bibnamefont {Gompper}},\ }\bibfield  {title} {\bibinfo {title} {Dynamics of vesicle self-assembly and dissolution},\ }\href@noop {} {\bibfield  {journal} {\bibinfo  {journal} {The Journal of chemical physics}\ }\textbf {\bibinfo {volume} {125}} (\bibinfo {year} {2006})}\BibitemShut {NoStop}%
\bibitem [{\citenamefont {Reynwar}\ \emph {et~al.}(2007)\citenamefont {Reynwar}, \citenamefont {Illya}, \citenamefont {Harmandaris}, \citenamefont {M{\"u}ller}, \citenamefont {Kremer},\ and\ \citenamefont {Deserno}}]{reynwar2007aggregation}%
  \BibitemOpen
  \bibfield  {author} {\bibinfo {author} {\bibfnamefont {B.~J.}\ \bibnamefont {Reynwar}}, \bibinfo {author} {\bibfnamefont {G.}~\bibnamefont {Illya}}, \bibinfo {author} {\bibfnamefont {V.~A.}\ \bibnamefont {Harmandaris}}, \bibinfo {author} {\bibfnamefont {M.~M.}\ \bibnamefont {M{\"u}ller}}, \bibinfo {author} {\bibfnamefont {K.}~\bibnamefont {Kremer}},\ and\ \bibinfo {author} {\bibfnamefont {M.}~\bibnamefont {Deserno}},\ }\bibfield  {title} {\bibinfo {title} {Aggregation and vesiculation of membrane proteins by curvature-mediated interactions},\ }\href@noop {} {\bibfield  {journal} {\bibinfo  {journal} {Nature}\ }\textbf {\bibinfo {volume} {447}},\ \bibinfo {pages} {461} (\bibinfo {year} {2007})}\BibitemShut {NoStop}%
\bibitem [{\citenamefont {Hu}\ \emph {et~al.}(2012)\citenamefont {Hu}, \citenamefont {Briguglio},\ and\ \citenamefont {Deserno}}]{Hu2012}%
  \BibitemOpen
  \bibfield  {author} {\bibinfo {author} {\bibfnamefont {M.}~\bibnamefont {Hu}}, \bibinfo {author} {\bibfnamefont {J.~J.}\ \bibnamefont {Briguglio}},\ and\ \bibinfo {author} {\bibfnamefont {M.}~\bibnamefont {Deserno}},\ }\bibfield  {title} {\bibinfo {title} {Determining the {Gaussian} curvature modulus of lipid membranes in simulations},\ }\bibfield  {journal} {\bibinfo  {journal} {Biophysical Journal}\ }\textbf {\bibinfo {volume} {102}},\ \href {https://doi.org/10.1016/j.bpj.2012.02.013} {10.1016/j.bpj.2012.02.013} (\bibinfo {year} {2012})\BibitemShut {NoStop}%
\bibitem [{\citenamefont {Asakura}\ and\ \citenamefont {Oosawa}(1954)}]{asakura1954interaction}%
  \BibitemOpen
  \bibfield  {author} {\bibinfo {author} {\bibfnamefont {S.}~\bibnamefont {Asakura}}\ and\ \bibinfo {author} {\bibfnamefont {F.}~\bibnamefont {Oosawa}},\ }\bibfield  {title} {\bibinfo {title} {On interaction between two bodies immersed in a solution of macromolecules},\ }\href@noop {} {\bibfield  {journal} {\bibinfo  {journal} {The Journal of chemical physics}\ }\textbf {\bibinfo {volume} {22}},\ \bibinfo {pages} {1255} (\bibinfo {year} {1954})}\BibitemShut {NoStop}%
\bibitem [{\citenamefont {Barry}\ and\ \citenamefont {Dogic}(2010)}]{Barry2010}%
  \BibitemOpen
  \bibfield  {author} {\bibinfo {author} {\bibfnamefont {E.}~\bibnamefont {Barry}}\ and\ \bibinfo {author} {\bibfnamefont {Z.}~\bibnamefont {Dogic}},\ }\bibfield  {title} {\bibinfo {title} {Entropy driven self-assembly of nonamphiphilic colloidal membranes},\ }\bibfield  {journal} {\bibinfo  {journal} {Proceedings of the National Academy of Sciences}\ }\textbf {\bibinfo {volume} {107}},\ \href {https://doi.org/10.1073/pnas.1000406107} {10.1073/pnas.1000406107} (\bibinfo {year} {2010})\BibitemShut {NoStop}%
\bibitem [{\citenamefont {Yang}\ \emph {et~al.}(2012)\citenamefont {Yang}, \citenamefont {Barry}, \citenamefont {Dogic},\ and\ \citenamefont {Hagan}}]{yang2012self}%
  \BibitemOpen
  \bibfield  {author} {\bibinfo {author} {\bibfnamefont {Y.}~\bibnamefont {Yang}}, \bibinfo {author} {\bibfnamefont {E.}~\bibnamefont {Barry}}, \bibinfo {author} {\bibfnamefont {Z.}~\bibnamefont {Dogic}},\ and\ \bibinfo {author} {\bibfnamefont {M.~F.}\ \bibnamefont {Hagan}},\ }\bibfield  {title} {\bibinfo {title} {Self-assembly of 2d membranes from mixtures of hard rods and depleting polymers},\ }\href@noop {} {\bibfield  {journal} {\bibinfo  {journal} {Soft Matter}\ }\textbf {\bibinfo {volume} {8}},\ \bibinfo {pages} {707} (\bibinfo {year} {2012})}\BibitemShut {NoStop}%
\bibitem [{\citenamefont {Kang}\ \emph {et~al.}(2016)\citenamefont {Kang}, \citenamefont {Gibaud}, \citenamefont {Dogic},\ and\ \citenamefont {Lubensky}}]{kang2016entropic}%
  \BibitemOpen
  \bibfield  {author} {\bibinfo {author} {\bibfnamefont {L.}~\bibnamefont {Kang}}, \bibinfo {author} {\bibfnamefont {T.}~\bibnamefont {Gibaud}}, \bibinfo {author} {\bibfnamefont {Z.}~\bibnamefont {Dogic}},\ and\ \bibinfo {author} {\bibfnamefont {T.}~\bibnamefont {Lubensky}},\ }\bibfield  {title} {\bibinfo {title} {Entropic forces stabilize diverse emergent structures in colloidal membranes},\ }\href@noop {} {\bibfield  {journal} {\bibinfo  {journal} {Soft matter}\ }\textbf {\bibinfo {volume} {12}},\ \bibinfo {pages} {386} (\bibinfo {year} {2016})}\BibitemShut {NoStop}%
\bibitem [{\citenamefont {Helfrich}(1973)}]{Helfrich1973}%
  \BibitemOpen
  \bibfield  {author} {\bibinfo {author} {\bibfnamefont {W.}~\bibnamefont {Helfrich}},\ }\bibfield  {title} {\bibinfo {title} {Elastic properties of lipid bilayers: Theory and possible experiments},\ }\bibfield  {journal} {\bibinfo  {journal} {Zeitschrift fur Naturforschung - Section C Journal of Biosciences}\ }\textbf {\bibinfo {volume} {28}},\ \href {https://doi.org/10.1515/znc-1973-11-1209} {10.1515/znc-1973-11-1209} (\bibinfo {year} {1973})\BibitemShut {NoStop}%
\bibitem [{\citenamefont {Faizi}\ \emph {et~al.}(2020)\citenamefont {Faizi}, \citenamefont {Reeves}, \citenamefont {Georgiev}, \citenamefont {Vlahovska},\ and\ \citenamefont {Dimova}}]{faizi2020fluctuation}%
  \BibitemOpen
  \bibfield  {author} {\bibinfo {author} {\bibfnamefont {H.~A.}\ \bibnamefont {Faizi}}, \bibinfo {author} {\bibfnamefont {C.~J.}\ \bibnamefont {Reeves}}, \bibinfo {author} {\bibfnamefont {V.~N.}\ \bibnamefont {Georgiev}}, \bibinfo {author} {\bibfnamefont {P.~M.}\ \bibnamefont {Vlahovska}},\ and\ \bibinfo {author} {\bibfnamefont {R.}~\bibnamefont {Dimova}},\ }\bibfield  {title} {\bibinfo {title} {Fluctuation spectroscopy of giant unilamellar vesicles using confocal and phase contrast microscopy},\ }\href@noop {} {\bibfield  {journal} {\bibinfo  {journal} {Soft Matter}\ }\textbf {\bibinfo {volume} {16}},\ \bibinfo {pages} {8996} (\bibinfo {year} {2020})}\BibitemShut {NoStop}%
\bibitem [{\citenamefont {Helfrich}(1974)}]{helfrich1974size}%
  \BibitemOpen
  \bibfield  {author} {\bibinfo {author} {\bibfnamefont {W.}~\bibnamefont {Helfrich}},\ }\bibfield  {title} {\bibinfo {title} {The size of bilayer vesicles generated by sonication},\ }\href@noop {} {\bibfield  {journal} {\bibinfo  {journal} {Physics letters A}\ }\textbf {\bibinfo {volume} {50}},\ \bibinfo {pages} {115} (\bibinfo {year} {1974})}\BibitemShut {NoStop}%
\bibitem [{\citenamefont {Fromherz}(1983)}]{fromherz1983lipid}%
  \BibitemOpen
  \bibfield  {author} {\bibinfo {author} {\bibfnamefont {P.}~\bibnamefont {Fromherz}},\ }\bibfield  {title} {\bibinfo {title} {Lipid-vesicle structure: size control by edge-active agents},\ }\href@noop {} {\bibfield  {journal} {\bibinfo  {journal} {Chemical physics letters}\ }\textbf {\bibinfo {volume} {94}},\ \bibinfo {pages} {259} (\bibinfo {year} {1983})}\BibitemShut {NoStop}%
\bibitem [{\citenamefont {Portet}\ and\ \citenamefont {Dimova}(2010)}]{portet2010new}%
  \BibitemOpen
  \bibfield  {author} {\bibinfo {author} {\bibfnamefont {T.}~\bibnamefont {Portet}}\ and\ \bibinfo {author} {\bibfnamefont {R.}~\bibnamefont {Dimova}},\ }\bibfield  {title} {\bibinfo {title} {A new method for measuring edge tensions and stability of lipid bilayers: effect of membrane composition},\ }\href@noop {} {\bibfield  {journal} {\bibinfo  {journal} {Biophysical journal}\ }\textbf {\bibinfo {volume} {99}},\ \bibinfo {pages} {3264} (\bibinfo {year} {2010})}\BibitemShut {NoStop}%
\bibitem [{\citenamefont {Szleifer}\ \emph {et~al.}(1988)\citenamefont {Szleifer}, \citenamefont {Kramer}, \citenamefont {Ben-Shaul}, \citenamefont {Roux},\ and\ \citenamefont {Gelbart}}]{Szleifer1988}%
  \BibitemOpen
  \bibfield  {author} {\bibinfo {author} {\bibfnamefont {I.}~\bibnamefont {Szleifer}}, \bibinfo {author} {\bibfnamefont {D.}~\bibnamefont {Kramer}}, \bibinfo {author} {\bibfnamefont {A.}~\bibnamefont {Ben-Shaul}}, \bibinfo {author} {\bibfnamefont {D.}~\bibnamefont {Roux}},\ and\ \bibinfo {author} {\bibfnamefont {W.~M.}\ \bibnamefont {Gelbart}},\ }\bibfield  {title} {\bibinfo {title} {Curvature elasticity of pure and mixed surfactant films.},\ }\href@noop {} {\bibfield  {journal} {\bibinfo  {journal} {Physical review letters}\ }\textbf {\bibinfo {volume} {60}} (\bibinfo {year} {1988})}\BibitemShut {NoStop}%
\bibitem [{\citenamefont {Rawicz}\ \emph {et~al.}(2000)\citenamefont {Rawicz}, \citenamefont {Olbrich}, \citenamefont {McIntosh}, \citenamefont {Needham},\ and\ \citenamefont {Evans}}]{rawicz2000effect}%
  \BibitemOpen
  \bibfield  {author} {\bibinfo {author} {\bibfnamefont {W.}~\bibnamefont {Rawicz}}, \bibinfo {author} {\bibfnamefont {K.~C.}\ \bibnamefont {Olbrich}}, \bibinfo {author} {\bibfnamefont {T.}~\bibnamefont {McIntosh}}, \bibinfo {author} {\bibfnamefont {D.}~\bibnamefont {Needham}},\ and\ \bibinfo {author} {\bibfnamefont {E.}~\bibnamefont {Evans}},\ }\bibfield  {title} {\bibinfo {title} {Effect of chain length and unsaturation on elasticity of lipid bilayers},\ }\href@noop {} {\bibfield  {journal} {\bibinfo  {journal} {Biophysical journal}\ }\textbf {\bibinfo {volume} {79}},\ \bibinfo {pages} {328} (\bibinfo {year} {2000})}\BibitemShut {NoStop}%
\bibitem [{\citenamefont {Balchunas}\ \emph {et~al.}(2019)\citenamefont {Balchunas}, \citenamefont {Cabanas}, \citenamefont {Zakhary}, \citenamefont {Gibaud}, \citenamefont {Fraden}, \citenamefont {Sharma}, \citenamefont {Hagan},\ and\ \citenamefont {Dogic}}]{Balchunas2019}%
  \BibitemOpen
  \bibfield  {author} {\bibinfo {author} {\bibfnamefont {A.~J.}\ \bibnamefont {Balchunas}}, \bibinfo {author} {\bibfnamefont {R.~A.}\ \bibnamefont {Cabanas}}, \bibinfo {author} {\bibfnamefont {M.~J.}\ \bibnamefont {Zakhary}}, \bibinfo {author} {\bibfnamefont {T.}~\bibnamefont {Gibaud}}, \bibinfo {author} {\bibfnamefont {S.}~\bibnamefont {Fraden}}, \bibinfo {author} {\bibfnamefont {P.}~\bibnamefont {Sharma}}, \bibinfo {author} {\bibfnamefont {M.~F.}\ \bibnamefont {Hagan}},\ and\ \bibinfo {author} {\bibfnamefont {Z.}~\bibnamefont {Dogic}},\ }\bibfield  {title} {\bibinfo {title} {Equation of state of colloidal membranes},\ }\bibfield  {journal} {\bibinfo  {journal} {Soft Matter}\ }\textbf {\bibinfo {volume} {15}},\ \href {https://doi.org/10.1039/c9sm01054h} {10.1039/c9sm01054h} (\bibinfo {year} {2019})\BibitemShut {NoStop}%
\bibitem [{\citenamefont {Zakhary}\ \emph {et~al.}(2014)\citenamefont {Zakhary}, \citenamefont {Gibaud}, \citenamefont {Kaplan}, \citenamefont {Barry}, \citenamefont {Oldenbourg}, \citenamefont {Meyer},\ and\ \citenamefont {Dogic}}]{Zakhary2014}%
  \BibitemOpen
  \bibfield  {author} {\bibinfo {author} {\bibfnamefont {M.~J.}\ \bibnamefont {Zakhary}}, \bibinfo {author} {\bibfnamefont {T.}~\bibnamefont {Gibaud}}, \bibinfo {author} {\bibfnamefont {C.~N.}\ \bibnamefont {Kaplan}}, \bibinfo {author} {\bibfnamefont {E.}~\bibnamefont {Barry}}, \bibinfo {author} {\bibfnamefont {R.}~\bibnamefont {Oldenbourg}}, \bibinfo {author} {\bibfnamefont {R.~B.}\ \bibnamefont {Meyer}},\ and\ \bibinfo {author} {\bibfnamefont {Z.}~\bibnamefont {Dogic}},\ }\bibfield  {title} {\bibinfo {title} {Imprintable membranes from incomplete chiral coalescence},\ }\bibfield  {journal} {\bibinfo  {journal} {Nature Communications}\ }\href {https://doi.org/10.1038/ncomms4063} {10.1038/ncomms4063} (\bibinfo {year} {2014})\BibitemShut {NoStop}%
\bibitem [{\citenamefont {Kraus}\ \emph {et~al.}(1995)\citenamefont {Kraus}, \citenamefont {Seifert},\ and\ \citenamefont {Lipowsky}}]{kraus1995gravity}%
  \BibitemOpen
  \bibfield  {author} {\bibinfo {author} {\bibfnamefont {M.}~\bibnamefont {Kraus}}, \bibinfo {author} {\bibfnamefont {U.}~\bibnamefont {Seifert}},\ and\ \bibinfo {author} {\bibfnamefont {R.}~\bibnamefont {Lipowsky}},\ }\bibfield  {title} {\bibinfo {title} {Gravity-induced shape transformations of vesicles},\ }\href@noop {} {\bibfield  {journal} {\bibinfo  {journal} {Europhyics letters}\ }\textbf {\bibinfo {volume} {32}},\ \bibinfo {pages} {431} (\bibinfo {year} {1995})}\BibitemShut {NoStop}%
\bibitem [{\citenamefont {Balchunas}\ \emph {et~al.}(2020)\citenamefont {Balchunas}, \citenamefont {Jia}, \citenamefont {Zakhary}, \citenamefont {Robaszewski}, \citenamefont {Gibaud}, \citenamefont {Dogic}, \citenamefont {Pelcovits},\ and\ \citenamefont {Powers}}]{Balchunas2020}%
  \BibitemOpen
  \bibfield  {author} {\bibinfo {author} {\bibfnamefont {A.}~\bibnamefont {Balchunas}}, \bibinfo {author} {\bibfnamefont {L.~L.}\ \bibnamefont {Jia}}, \bibinfo {author} {\bibfnamefont {M.~J.}\ \bibnamefont {Zakhary}}, \bibinfo {author} {\bibfnamefont {J.}~\bibnamefont {Robaszewski}}, \bibinfo {author} {\bibfnamefont {T.}~\bibnamefont {Gibaud}}, \bibinfo {author} {\bibfnamefont {Z.}~\bibnamefont {Dogic}}, \bibinfo {author} {\bibfnamefont {R.~A.}\ \bibnamefont {Pelcovits}},\ and\ \bibinfo {author} {\bibfnamefont {T.~R.}\ \bibnamefont {Powers}},\ }\bibfield  {title} {\bibinfo {title} {Force-induced formation of twisted chiral ribbons},\ }\bibfield  {journal} {\bibinfo  {journal} {Physical Review Letters}\ }\textbf {\bibinfo {volume} {125}},\ \href {https://doi.org/10.1103/PhysRevLett.125.018002} {10.1103/PhysRevLett.125.018002} (\bibinfo {year} {2020})\BibitemShut {NoStop}%
\bibitem [{\citenamefont {Khanra}\ \emph {et~al.}(2022)\citenamefont {Khanra}, \citenamefont {Jia}, \citenamefont {Mitchell}, \citenamefont {Balchunas}, \citenamefont {Pelcovits}, \citenamefont {Powers}, \citenamefont {Dogic},\ and\ \citenamefont {Sharma}}]{Khanra2022}%
  \BibitemOpen
  \bibfield  {author} {\bibinfo {author} {\bibfnamefont {A.}~\bibnamefont {Khanra}}, \bibinfo {author} {\bibfnamefont {L.~L.}\ \bibnamefont {Jia}}, \bibinfo {author} {\bibfnamefont {N.~P.}\ \bibnamefont {Mitchell}}, \bibinfo {author} {\bibfnamefont {A.}~\bibnamefont {Balchunas}}, \bibinfo {author} {\bibfnamefont {R.~A.}\ \bibnamefont {Pelcovits}}, \bibinfo {author} {\bibfnamefont {T.~R.}\ \bibnamefont {Powers}}, \bibinfo {author} {\bibfnamefont {Z.}~\bibnamefont {Dogic}},\ and\ \bibinfo {author} {\bibfnamefont {P.}~\bibnamefont {Sharma}},\ }\bibfield  {title} {\bibinfo {title} {Controlling the shape and topology of two-component colloidal membranes},\ }\bibfield  {journal} {\bibinfo  {journal} {Proceedings of the National Academy of Sciences}\ }\textbf {\bibinfo {volume} {119}},\ \href {https://doi.org/10.1073/pnas.2204453119} {10.1073/pnas.2204453119} (\bibinfo {year} {2022})\BibitemShut {NoStop}%
\bibitem [{\citenamefont {Powers}\ \emph {et~al.}(2002)\citenamefont {Powers}, \citenamefont {Huber},\ and\ \citenamefont {Goldstein}}]{powers2002fluid}%
  \BibitemOpen
  \bibfield  {author} {\bibinfo {author} {\bibfnamefont {T.~R.}\ \bibnamefont {Powers}}, \bibinfo {author} {\bibfnamefont {G.}~\bibnamefont {Huber}},\ and\ \bibinfo {author} {\bibfnamefont {R.~E.}\ \bibnamefont {Goldstein}},\ }\bibfield  {title} {\bibinfo {title} {Fluid-membrane tethers: minimal surfaces and elastic boundary layers},\ }\href@noop {} {\bibfield  {journal} {\bibinfo  {journal} {Physical review E}\ }\textbf {\bibinfo {volume} {65}},\ \bibinfo {pages} {041901} (\bibinfo {year} {2002})}\BibitemShut {NoStop}%
\bibitem [{\citenamefont {Roux}\ \emph {et~al.}(2002)\citenamefont {Roux}, \citenamefont {Cappello}, \citenamefont {Cartaud}, \citenamefont {Prost}, \citenamefont {Goud},\ and\ \citenamefont {Bassereau}}]{roux2002minimal}%
  \BibitemOpen
  \bibfield  {author} {\bibinfo {author} {\bibfnamefont {A.}~\bibnamefont {Roux}}, \bibinfo {author} {\bibfnamefont {G.}~\bibnamefont {Cappello}}, \bibinfo {author} {\bibfnamefont {J.}~\bibnamefont {Cartaud}}, \bibinfo {author} {\bibfnamefont {J.}~\bibnamefont {Prost}}, \bibinfo {author} {\bibfnamefont {B.}~\bibnamefont {Goud}},\ and\ \bibinfo {author} {\bibfnamefont {P.}~\bibnamefont {Bassereau}},\ }\bibfield  {title} {\bibinfo {title} {A minimal system allowing tubulation with molecular motors pulling on giant liposomes},\ }\href@noop {} {\bibfield  {journal} {\bibinfo  {journal} {Proceedings of the National Academy of Sciences}\ }\textbf {\bibinfo {volume} {99}},\ \bibinfo {pages} {5394} (\bibinfo {year} {2002})}\BibitemShut {NoStop}%
\bibitem [{\citenamefont {Karatekin}\ \emph {et~al.}(2003)\citenamefont {Karatekin}, \citenamefont {Sandre}, \citenamefont {Guitouni}, \citenamefont {Borghi}, \citenamefont {Puech},\ and\ \citenamefont {Brochard-Wyart}}]{karatekin2003cascades}%
  \BibitemOpen
  \bibfield  {author} {\bibinfo {author} {\bibfnamefont {E.}~\bibnamefont {Karatekin}}, \bibinfo {author} {\bibfnamefont {O.}~\bibnamefont {Sandre}}, \bibinfo {author} {\bibfnamefont {H.}~\bibnamefont {Guitouni}}, \bibinfo {author} {\bibfnamefont {N.}~\bibnamefont {Borghi}}, \bibinfo {author} {\bibfnamefont {P.-H.}\ \bibnamefont {Puech}},\ and\ \bibinfo {author} {\bibfnamefont {F.}~\bibnamefont {Brochard-Wyart}},\ }\bibfield  {title} {\bibinfo {title} {Cascades of transient pores in giant vesicles: line tension and transport},\ }\href@noop {} {\bibfield  {journal} {\bibinfo  {journal} {Biophysical journal}\ }\textbf {\bibinfo {volume} {84}},\ \bibinfo {pages} {1734} (\bibinfo {year} {2003})}\BibitemShut {NoStop}%
\bibitem [{\citenamefont {Oglecka}\ \emph {et~al.}(2014)\citenamefont {Oglecka}, \citenamefont {Rangamani}, \citenamefont {Liedberg}, \citenamefont {Kraut},\ and\ \citenamefont {Parikh}}]{oglkecka2014oscillatory}%
  \BibitemOpen
  \bibfield  {author} {\bibinfo {author} {\bibfnamefont {K.}~\bibnamefont {Oglecka}}, \bibinfo {author} {\bibfnamefont {P.}~\bibnamefont {Rangamani}}, \bibinfo {author} {\bibfnamefont {B.}~\bibnamefont {Liedberg}}, \bibinfo {author} {\bibfnamefont {R.~S.}\ \bibnamefont {Kraut}},\ and\ \bibinfo {author} {\bibfnamefont {A.~N.}\ \bibnamefont {Parikh}},\ }\bibfield  {title} {\bibinfo {title} {Oscillatory phase separation in giant lipid vesicles induced by transmembrane osmotic differentials},\ }\href@noop {} {\bibfield  {journal} {\bibinfo  {journal} {elife}\ }\textbf {\bibinfo {volume} {3}},\ \bibinfo {pages} {e03695} (\bibinfo {year} {2014})}\BibitemShut {NoStop}%
\bibitem [{\citenamefont {Sandre}\ \emph {et~al.}(1999)\citenamefont {Sandre}, \citenamefont {Moreaux},\ and\ \citenamefont {Brochard-Wyart}}]{Sandre1999}%
  \BibitemOpen
  \bibfield  {author} {\bibinfo {author} {\bibfnamefont {O.}~\bibnamefont {Sandre}}, \bibinfo {author} {\bibfnamefont {L.}~\bibnamefont {Moreaux}},\ and\ \bibinfo {author} {\bibfnamefont {F.}~\bibnamefont {Brochard-Wyart}},\ }\bibfield  {title} {\bibinfo {title} {Dynamics of transient pores in stretched vesicles},\ }\bibfield  {journal} {\bibinfo  {journal} {Proceedings of the National Academy of Sciences}\ }\href {https://doi.org/10.1073/pnas.96.19.10591} {10.1073/pnas.96.19.10591} (\bibinfo {year} {1999})\BibitemShut {NoStop}%
\bibitem [{\citenamefont {Chabanon}\ \emph {et~al.}(2017)\citenamefont {Chabanon}, \citenamefont {Ho}, \citenamefont {Liedberg}, \citenamefont {Parikh},\ and\ \citenamefont {Rangamani}}]{Chabanon2017}%
  \BibitemOpen
  \bibfield  {author} {\bibinfo {author} {\bibfnamefont {M.}~\bibnamefont {Chabanon}}, \bibinfo {author} {\bibfnamefont {J.~C.}\ \bibnamefont {Ho}}, \bibinfo {author} {\bibfnamefont {B.}~\bibnamefont {Liedberg}}, \bibinfo {author} {\bibfnamefont {A.~N.}\ \bibnamefont {Parikh}},\ and\ \bibinfo {author} {\bibfnamefont {P.}~\bibnamefont {Rangamani}},\ }\bibfield  {title} {\bibinfo {title} {Pulsatile lipid vesicles under osmotic stress},\ }\bibfield  {journal} {\bibinfo  {journal} {Biophysical Journal}\ }\textbf {\bibinfo {volume} {112}},\ \href {https://doi.org/10.1016/j.bpj.2017.03.018} {10.1016/j.bpj.2017.03.018} (\bibinfo {year} {2017})\BibitemShut {NoStop}%
\bibitem [{\citenamefont {Moroz}\ and\ \citenamefont {Nelson}(1997)}]{Moroz1997}%
  \BibitemOpen
  \bibfield  {author} {\bibinfo {author} {\bibfnamefont {J.}~\bibnamefont {Moroz}}\ and\ \bibinfo {author} {\bibfnamefont {P.}~\bibnamefont {Nelson}},\ }\bibfield  {title} {\bibinfo {title} {Dynamically stabilized pores in bilayer membranes},\ }\bibfield  {journal} {\bibinfo  {journal} {Biophysical Journal}\ }\href {https://doi.org/10.1016/S0006-3495(97)78864-7} {10.1016/S0006-3495(97)78864-7} (\bibinfo {year} {1997})\BibitemShut {NoStop}%
\bibitem [{\citenamefont {Dimova}\ \emph {et~al.}(2009)\citenamefont {Dimova}, \citenamefont {Bezlyepkina}, \citenamefont {Jord{\"o}}, \citenamefont {Knorr}, \citenamefont {Riske}, \citenamefont {Staykova}, \citenamefont {Vlahovska}, \citenamefont {Yamamoto}, \citenamefont {Yang},\ and\ \citenamefont {Lipowsky}}]{dimova2009vesicles}%
  \BibitemOpen
  \bibfield  {author} {\bibinfo {author} {\bibfnamefont {R.}~\bibnamefont {Dimova}}, \bibinfo {author} {\bibfnamefont {N.}~\bibnamefont {Bezlyepkina}}, \bibinfo {author} {\bibfnamefont {M.~D.}\ \bibnamefont {Jord{\"o}}}, \bibinfo {author} {\bibfnamefont {R.~L.}\ \bibnamefont {Knorr}}, \bibinfo {author} {\bibfnamefont {K.~A.}\ \bibnamefont {Riske}}, \bibinfo {author} {\bibfnamefont {M.}~\bibnamefont {Staykova}}, \bibinfo {author} {\bibfnamefont {P.~M.}\ \bibnamefont {Vlahovska}}, \bibinfo {author} {\bibfnamefont {T.}~\bibnamefont {Yamamoto}}, \bibinfo {author} {\bibfnamefont {P.}~\bibnamefont {Yang}},\ and\ \bibinfo {author} {\bibfnamefont {R.}~\bibnamefont {Lipowsky}},\ }\bibfield  {title} {\bibinfo {title} {Vesicles in electric fields: Some novel aspects of membrane behavior},\ }\href@noop {} {\bibfield  {journal} {\bibinfo  {journal} {Soft Matter}\ }\textbf {\bibinfo {volume} {5}},\ \bibinfo {pages} {3201} (\bibinfo {year} {2009})}\BibitemShut {NoStop}%
\bibitem [{\citenamefont {Hamada}\ \emph {et~al.}(2010)\citenamefont {Hamada}, \citenamefont {Sugimoto}, \citenamefont {Vestergaard}, \citenamefont {Nagasaki},\ and\ \citenamefont {Takagi}}]{hamada2010membrane}%
  \BibitemOpen
  \bibfield  {author} {\bibinfo {author} {\bibfnamefont {T.}~\bibnamefont {Hamada}}, \bibinfo {author} {\bibfnamefont {R.}~\bibnamefont {Sugimoto}}, \bibinfo {author} {\bibfnamefont {M.~C.}\ \bibnamefont {Vestergaard}}, \bibinfo {author} {\bibfnamefont {T.}~\bibnamefont {Nagasaki}},\ and\ \bibinfo {author} {\bibfnamefont {M.}~\bibnamefont {Takagi}},\ }\bibfield  {title} {\bibinfo {title} {Membrane disk and sphere: controllable mesoscopic structures for the capture and release of a targeted object},\ }\href@noop {} {\bibfield  {journal} {\bibinfo  {journal} {Journal of the American Chemical Society}\ }\textbf {\bibinfo {volume} {132}},\ \bibinfo {pages} {10528} (\bibinfo {year} {2010})}\BibitemShut {NoStop}%
\bibitem [{\citenamefont {Malik}\ \emph {et~al.}(2022)\citenamefont {Malik}, \citenamefont {Pak},\ and\ \citenamefont {Feng}}]{malik2022pore}%
  \BibitemOpen
  \bibfield  {author} {\bibinfo {author} {\bibfnamefont {V.~K.}\ \bibnamefont {Malik}}, \bibinfo {author} {\bibfnamefont {O.~S.}\ \bibnamefont {Pak}},\ and\ \bibinfo {author} {\bibfnamefont {J.}~\bibnamefont {Feng}},\ }\bibfield  {title} {\bibinfo {title} {Pore dynamics of lipid vesicles under light-induced osmotic stress},\ }\href@noop {} {\bibfield  {journal} {\bibinfo  {journal} {Physical Review Applied}\ }\textbf {\bibinfo {volume} {17}},\ \bibinfo {pages} {024032} (\bibinfo {year} {2022})}\BibitemShut {NoStop}%
\bibitem [{\citenamefont {Boal}\ and\ \citenamefont {Rao}(1992)}]{Boal1992}%
  \BibitemOpen
  \bibfield  {author} {\bibinfo {author} {\bibfnamefont {D.~H.}\ \bibnamefont {Boal}}\ and\ \bibinfo {author} {\bibfnamefont {M.}~\bibnamefont {Rao}},\ }\bibfield  {title} {\bibinfo {title} {Topology changes in fluid membranes.},\ }\href@noop {} {\bibfield  {journal} {\bibinfo  {journal} {Physical Review A}\ } (\bibinfo {year} {1992})}\BibitemShut {NoStop}%
\bibitem [{\citenamefont {Devanand}\ and\ \citenamefont {Selser}(1991)}]{Devanand1991}%
  \BibitemOpen
  \bibfield  {author} {\bibinfo {author} {\bibfnamefont {K.}~\bibnamefont {Devanand}}\ and\ \bibinfo {author} {\bibfnamefont {J.~C.}\ \bibnamefont {Selser}},\ }\bibfield  {title} {\bibinfo {title} {Asymptotic behavior and long-range interactions in aqueous solutions of poly(ethylene oxide)},\ }\href@noop {} {\bibfield  {journal} {\bibinfo  {journal} {Macromolecules}\ }\textbf {\bibinfo {volume} {24}},\ \bibinfo {pages} {5943} (\bibinfo {year} {1991})}\BibitemShut {NoStop}%
\bibitem [{\citenamefont {Zhang}\ \emph {et~al.}(2018)\citenamefont {Zhang}, \citenamefont {Zhang}, \citenamefont {Fang}, \citenamefont {Zhang}, \citenamefont {Wang},\ and\ \citenamefont {Jin}}]{Zhang2018}%
  \BibitemOpen
  \bibfield  {author} {\bibinfo {author} {\bibfnamefont {S.}~\bibnamefont {Zhang}}, \bibinfo {author} {\bibfnamefont {J.}~\bibnamefont {Zhang}}, \bibinfo {author} {\bibfnamefont {W.}~\bibnamefont {Fang}}, \bibinfo {author} {\bibfnamefont {Y.}~\bibnamefont {Zhang}}, \bibinfo {author} {\bibfnamefont {Q.}~\bibnamefont {Wang}},\ and\ \bibinfo {author} {\bibfnamefont {J.}~\bibnamefont {Jin}},\ }\bibfield  {title} {\bibinfo {title} {Ultralarge single-layer porous protein nanosheet for precise nanosize separation.},\ }\href@noop {} {\bibfield  {journal} {\bibinfo  {journal} {Nano Letters}\ }\textbf {\bibinfo {volume} {18}},\ \bibinfo {pages} {6563} (\bibinfo {year} {2018})}\BibitemShut {NoStop}%
\bibitem [{\citenamefont {Gibaud}\ \emph {et~al.}(2012{\natexlab{a}})\citenamefont {Gibaud}, \citenamefont {Barry}, \citenamefont {Zakhary}, \citenamefont {Henglin}, \citenamefont {Ward}, \citenamefont {Yang}, \citenamefont {Berciu}, \citenamefont {Oldenbourg}, \citenamefont {Hagan}, \citenamefont {Nicastro} \emph {et~al.}}]{gibaud2012reconfigurable}%
  \BibitemOpen
  \bibfield  {author} {\bibinfo {author} {\bibfnamefont {T.}~\bibnamefont {Gibaud}}, \bibinfo {author} {\bibfnamefont {E.}~\bibnamefont {Barry}}, \bibinfo {author} {\bibfnamefont {M.~J.}\ \bibnamefont {Zakhary}}, \bibinfo {author} {\bibfnamefont {M.}~\bibnamefont {Henglin}}, \bibinfo {author} {\bibfnamefont {A.}~\bibnamefont {Ward}}, \bibinfo {author} {\bibfnamefont {Y.}~\bibnamefont {Yang}}, \bibinfo {author} {\bibfnamefont {C.}~\bibnamefont {Berciu}}, \bibinfo {author} {\bibfnamefont {R.}~\bibnamefont {Oldenbourg}}, \bibinfo {author} {\bibfnamefont {M.~F.}\ \bibnamefont {Hagan}}, \bibinfo {author} {\bibfnamefont {D.}~\bibnamefont {Nicastro}}, \emph {et~al.},\ }\bibfield  {title} {\bibinfo {title} {Reconfigurable self-assembly through chiral control of interfacial tension},\ }\href@noop {} {\bibfield  {journal} {\bibinfo  {journal} {Nature}\ }\textbf {\bibinfo {volume} {481}},\ \bibinfo {pages} {348} (\bibinfo {year} {2012}{\natexlab{a}})}\BibitemShut {NoStop}%
\bibitem [{\citenamefont {Sharma}\ \emph {et~al.}(2014)\citenamefont {Sharma}, \citenamefont {Ward}, \citenamefont {Gibaud}, \citenamefont {Hagan},\ and\ \citenamefont {Dogic}}]{sharma2014hierarchical}%
  \BibitemOpen
  \bibfield  {author} {\bibinfo {author} {\bibfnamefont {P.}~\bibnamefont {Sharma}}, \bibinfo {author} {\bibfnamefont {A.}~\bibnamefont {Ward}}, \bibinfo {author} {\bibfnamefont {T.}~\bibnamefont {Gibaud}}, \bibinfo {author} {\bibfnamefont {M.~F.}\ \bibnamefont {Hagan}},\ and\ \bibinfo {author} {\bibfnamefont {Z.}~\bibnamefont {Dogic}},\ }\bibfield  {title} {\bibinfo {title} {Hierarchical organization of chiral rafts in colloidal membranes},\ }\href@noop {} {\bibfield  {journal} {\bibinfo  {journal} {Nature}\ }\textbf {\bibinfo {volume} {513}},\ \bibinfo {pages} {77} (\bibinfo {year} {2014})}\BibitemShut {NoStop}%
\bibitem [{\citenamefont {Gibaud}\ \emph {et~al.}(2017{\natexlab{a}})\citenamefont {Gibaud}, \citenamefont {Kaplan}, \citenamefont {Sharma}, \citenamefont {Zakhary}, \citenamefont {Ward}, \citenamefont {Oldenbourg}, \citenamefont {Meyer}, \citenamefont {Kamien}, \citenamefont {Powers},\ and\ \citenamefont {Dogic}}]{gibaud2017achiral}%
  \BibitemOpen
  \bibfield  {author} {\bibinfo {author} {\bibfnamefont {T.}~\bibnamefont {Gibaud}}, \bibinfo {author} {\bibfnamefont {C.~N.}\ \bibnamefont {Kaplan}}, \bibinfo {author} {\bibfnamefont {P.}~\bibnamefont {Sharma}}, \bibinfo {author} {\bibfnamefont {M.~J.}\ \bibnamefont {Zakhary}}, \bibinfo {author} {\bibfnamefont {A.}~\bibnamefont {Ward}}, \bibinfo {author} {\bibfnamefont {R.}~\bibnamefont {Oldenbourg}}, \bibinfo {author} {\bibfnamefont {R.~B.}\ \bibnamefont {Meyer}}, \bibinfo {author} {\bibfnamefont {R.~D.}\ \bibnamefont {Kamien}}, \bibinfo {author} {\bibfnamefont {T.~R.}\ \bibnamefont {Powers}},\ and\ \bibinfo {author} {\bibfnamefont {Z.}~\bibnamefont {Dogic}},\ }\bibfield  {title} {\bibinfo {title} {Achiral symmetry breaking and positive gaussian modulus lead to scalloped colloidal membranes},\ }\href@noop {} {\bibfield  {journal} {\bibinfo  {journal} {Proceedings of the National Academy of Sciences}\ }\textbf {\bibinfo {volume} {114}},\ \bibinfo {pages} {E3376} (\bibinfo {year}
  {2017}{\natexlab{a}})}\BibitemShut {NoStop}%
\bibitem [{\citenamefont {Wood}(1983)}]{Wood1983}%
  \BibitemOpen
  \bibfield  {author} {\bibinfo {author} {\bibfnamefont {E.}~\bibnamefont {Wood}},\ }\bibfield  {title} {\bibinfo {title} {Molecular cloning. {A} laboratory manual},\ }\bibfield  {journal} {\bibinfo  {journal} {Biochemical Education}\ }\textbf {\bibinfo {volume} {11}},\ \href {https://doi.org/10.1016/0307-4412(83)90068-7} {10.1016/0307-4412(83)90068-7} (\bibinfo {year} {1983})\BibitemShut {NoStop}%
\bibitem [{\citenamefont {Evans}\ \emph {et~al.}(1995)\citenamefont {Evans}, \citenamefont {Cook}, \citenamefont {Riggs},\ and\ \citenamefont {Noren}}]{evans1995litmus}%
  \BibitemOpen
  \bibfield  {author} {\bibinfo {author} {\bibfnamefont {P.}~\bibnamefont {Evans}}, \bibinfo {author} {\bibfnamefont {S.}~\bibnamefont {Cook}}, \bibinfo {author} {\bibfnamefont {P.}~\bibnamefont {Riggs}},\ and\ \bibinfo {author} {\bibfnamefont {C.}~\bibnamefont {Noren}},\ }\bibfield  {title} {\bibinfo {title} {Litmus: multipurpose cloning vectors with a novel system for bidirectional in vitro transcription},\ }\href@noop {} {\bibfield  {journal} {\bibinfo  {journal} {Biotechniques}\ }\textbf {\bibinfo {volume} {19}},\ \bibinfo {pages} {130} (\bibinfo {year} {1995})}\BibitemShut {NoStop}%
\bibitem [{\citenamefont {Monjezi}\ \emph {et~al.}(2010)\citenamefont {Monjezi}, \citenamefont {Tey}, \citenamefont {Sieo},\ and\ \citenamefont {Tan}}]{Monjezi2010}%
  \BibitemOpen
  \bibfield  {author} {\bibinfo {author} {\bibfnamefont {R.}~\bibnamefont {Monjezi}}, \bibinfo {author} {\bibfnamefont {B.~T.}\ \bibnamefont {Tey}}, \bibinfo {author} {\bibfnamefont {C.~C.}\ \bibnamefont {Sieo}},\ and\ \bibinfo {author} {\bibfnamefont {W.~S.}\ \bibnamefont {Tan}},\ }\bibfield  {title} {\bibinfo {title} {Purification of bacteriophage {M13} by anion exchange chromatography},\ }\bibfield  {journal} {\bibinfo  {journal} {Journal of Chromatography B}\ }\textbf {\bibinfo {volume} {878}},\ \href {https://doi.org/10.1016/j.jchromb.2010.05.028} {10.1016/j.jchromb.2010.05.028} (\bibinfo {year} {2010})\BibitemShut {NoStop}%
\bibitem [{\citenamefont {Lau}\ \emph {et~al.}(2009)\citenamefont {Lau}, \citenamefont {Prasad},\ and\ \citenamefont {Dogic}}]{Lau2009}%
  \BibitemOpen
  \bibfield  {author} {\bibinfo {author} {\bibfnamefont {A.~W.~C.}\ \bibnamefont {Lau}}, \bibinfo {author} {\bibfnamefont {A.}~\bibnamefont {Prasad}},\ and\ \bibinfo {author} {\bibfnamefont {Z.}~\bibnamefont {Dogic}},\ }\bibfield  {title} {\bibinfo {title} {Condensation of isolated semi-flexible filaments driven by depletion interactions},\ }\bibfield  {journal} {\bibinfo  {journal} {Europhysics Letters}\ }\textbf {\bibinfo {volume} {87}},\ \href {https://doi.org/10.1209/0295-5075/87/48006} {10.1209/0295-5075/87/48006} (\bibinfo {year} {2009})\BibitemShut {NoStop}%
\bibitem [{\citenamefont {Praetorius}\ \emph {et~al.}(2017)\citenamefont {Praetorius}, \citenamefont {Kick}, \citenamefont {Behler}, \citenamefont {Honemann}, \citenamefont {Weuster-Botz},\ and\ \citenamefont {Dietz}}]{pScaf2017}%
  \BibitemOpen
  \bibfield  {author} {\bibinfo {author} {\bibfnamefont {F.}~\bibnamefont {Praetorius}}, \bibinfo {author} {\bibfnamefont {B.}~\bibnamefont {Kick}}, \bibinfo {author} {\bibfnamefont {K.~L.}\ \bibnamefont {Behler}}, \bibinfo {author} {\bibfnamefont {M.~N.}\ \bibnamefont {Honemann}}, \bibinfo {author} {\bibfnamefont {D.}~\bibnamefont {Weuster-Botz}},\ and\ \bibinfo {author} {\bibfnamefont {H.}~\bibnamefont {Dietz}},\ }\bibfield  {title} {\bibinfo {title} {Biotechnological mass production of {DNA} origami.},\ }\href@noop {} {\bibfield  {journal} {\bibinfo  {journal} {Nature}\ }\textbf {\bibinfo {volume} {552}},\ \bibinfo {pages} {84} (\bibinfo {year} {2017})}\BibitemShut {NoStop}%
\bibitem [{\citenamefont {Nafisi}\ \emph {et~al.}(2018)\citenamefont {Nafisi}, \citenamefont {Aksel},\ and\ \citenamefont {Douglas}}]{pScaf2018}%
  \BibitemOpen
  \bibfield  {author} {\bibinfo {author} {\bibfnamefont {P.~M.}\ \bibnamefont {Nafisi}}, \bibinfo {author} {\bibfnamefont {T.}~\bibnamefont {Aksel}},\ and\ \bibinfo {author} {\bibfnamefont {S.~M.}\ \bibnamefont {Douglas}},\ }\bibfield  {title} {\bibinfo {title} {Construction of a novel phagemid to produce custom {DNA} origami scaffolds.},\ }\href@noop {} {\bibfield  {journal} {\bibinfo  {journal} {Synthetic Biology}\ }\textbf {\bibinfo {volume} {3}} (\bibinfo {year} {2018})}\BibitemShut {NoStop}%
\bibitem [{\citenamefont {Schindelin}\ \emph {et~al.}(2012)\citenamefont {Schindelin}, \citenamefont {Arganda-Carreras}, \citenamefont {Frise}, \citenamefont {Kaynig}, \citenamefont {Longair}, \citenamefont {Pietzsch}, \citenamefont {Preibisch}, \citenamefont {Rueden}, \citenamefont {Saalfeld}, \citenamefont {Schmid}, \citenamefont {Tinevez}, \citenamefont {White}, \citenamefont {Hartenstein}, \citenamefont {Eliceiri}, \citenamefont {Tomancak},\ and\ \citenamefont {Cardona}}]{Fiji2012}%
  \BibitemOpen
  \bibfield  {author} {\bibinfo {author} {\bibfnamefont {J.}~\bibnamefont {Schindelin}}, \bibinfo {author} {\bibfnamefont {I.}~\bibnamefont {Arganda-Carreras}}, \bibinfo {author} {\bibfnamefont {E.}~\bibnamefont {Frise}}, \bibinfo {author} {\bibfnamefont {V.}~\bibnamefont {Kaynig}}, \bibinfo {author} {\bibfnamefont {M.}~\bibnamefont {Longair}}, \bibinfo {author} {\bibfnamefont {T.}~\bibnamefont {Pietzsch}}, \bibinfo {author} {\bibfnamefont {S.}~\bibnamefont {Preibisch}}, \bibinfo {author} {\bibfnamefont {C.}~\bibnamefont {Rueden}}, \bibinfo {author} {\bibfnamefont {S.}~\bibnamefont {Saalfeld}}, \bibinfo {author} {\bibfnamefont {B.}~\bibnamefont {Schmid}}, \bibinfo {author} {\bibfnamefont {J.-Y.}\ \bibnamefont {Tinevez}}, \bibinfo {author} {\bibfnamefont {D.~J.}\ \bibnamefont {White}}, \bibinfo {author} {\bibfnamefont {V.}~\bibnamefont {Hartenstein}}, \bibinfo {author} {\bibfnamefont {K.}~\bibnamefont {Eliceiri}}, \bibinfo {author} {\bibfnamefont {P.}~\bibnamefont {Tomancak}},\ and\ \bibinfo {author}
  {\bibfnamefont {A.}~\bibnamefont {Cardona}},\ }\bibfield  {title} {\bibinfo {title} {Fiji: an open-source platform for biological-image analysis},\ }\href@noop {} {\bibfield  {journal} {\bibinfo  {journal} {Nature Methods}\ }\textbf {\bibinfo {volume} {9}},\ \bibinfo {pages} {676} (\bibinfo {year} {2012})}\BibitemShut {NoStop}%
\bibitem [{\citenamefont {Steger}(1998)}]{Steger1998}%
  \BibitemOpen
  \bibfield  {author} {\bibinfo {author} {\bibfnamefont {C.}~\bibnamefont {Steger}},\ }\bibfield  {title} {\bibinfo {title} {An unbiased detector of curvilinear structures.},\ }\bibfield  {journal} {\bibinfo  {journal} {EEE Transactions on Pattern Analysis and Machine Intelligence}\ }\textbf {\bibinfo {volume} {20}},\ \href {https://doi.org/10.1109/34.659930} {10.1109/34.659930} (\bibinfo {year} {1998})\BibitemShut {NoStop}%
\bibitem [{\citenamefont {Cignoni}\ \emph {et~al.}(2008)\citenamefont {Cignoni}, \citenamefont {Callieri}, \citenamefont {Corsini}, \citenamefont {Dellepiane}, \citenamefont {Ganovelli},\ and\ \citenamefont {Ranzuglia}}]{MeshLab}%
  \BibitemOpen
  \bibfield  {author} {\bibinfo {author} {\bibfnamefont {P.}~\bibnamefont {Cignoni}}, \bibinfo {author} {\bibfnamefont {M.}~\bibnamefont {Callieri}}, \bibinfo {author} {\bibfnamefont {M.}~\bibnamefont {Corsini}}, \bibinfo {author} {\bibfnamefont {M.}~\bibnamefont {Dellepiane}}, \bibinfo {author} {\bibfnamefont {F.}~\bibnamefont {Ganovelli}},\ and\ \bibinfo {author} {\bibfnamefont {G.}~\bibnamefont {Ranzuglia}},\ }\bibfield  {title} {\bibinfo {title} {Meshlab: an open-source mesh processing tool.},\ }\href@noop {} {\bibfield  {journal} {\bibinfo  {journal} {Eurographics Italian chapter conference}\ } (\bibinfo {year} {2008})}\BibitemShut {NoStop}%
\bibitem [{\citenamefont {Mutz}\ and\ \citenamefont {Helfrich}(1990)}]{Mutz1990}%
  \BibitemOpen
  \bibfield  {author} {\bibinfo {author} {\bibfnamefont {M.}~\bibnamefont {Mutz}}\ and\ \bibinfo {author} {\bibfnamefont {W.}~\bibnamefont {Helfrich}},\ }\bibfield  {title} {\bibinfo {title} {Bending rigidities of some biological model membranes as obtained from the {F}ourier analysis of contour sections},\ }\bibfield  {journal} {\bibinfo  {journal} {Journal de Physique}\ }\textbf {\bibinfo {volume} {51}},\ \href {https://doi.org/10.1051/jphys:019900051010099100} {10.1051/jphys:019900051010099100} (\bibinfo {year} {1990})\BibitemShut {NoStop}%
\bibitem [{\citenamefont {Gibaud}\ \emph {et~al.}(2012{\natexlab{b}})\citenamefont {Gibaud}, \citenamefont {Barry}, \citenamefont {Zakhary}, \citenamefont {Henglin}, \citenamefont {Ward}, \citenamefont {Yang}, \citenamefont {Berciu}, \citenamefont {Oldenbourg}, \citenamefont {Hagan}, \citenamefont {Nicastro}, \citenamefont {Meyer},\ and\ \citenamefont {Dogic}}]{Gibaud2012}%
  \BibitemOpen
  \bibfield  {author} {\bibinfo {author} {\bibfnamefont {T.}~\bibnamefont {Gibaud}}, \bibinfo {author} {\bibfnamefont {E.}~\bibnamefont {Barry}}, \bibinfo {author} {\bibfnamefont {M.~J.}\ \bibnamefont {Zakhary}}, \bibinfo {author} {\bibfnamefont {M.}~\bibnamefont {Henglin}}, \bibinfo {author} {\bibfnamefont {A.}~\bibnamefont {Ward}}, \bibinfo {author} {\bibfnamefont {Y.}~\bibnamefont {Yang}}, \bibinfo {author} {\bibfnamefont {C.}~\bibnamefont {Berciu}}, \bibinfo {author} {\bibfnamefont {R.}~\bibnamefont {Oldenbourg}}, \bibinfo {author} {\bibfnamefont {M.~F.}\ \bibnamefont {Hagan}}, \bibinfo {author} {\bibfnamefont {D.}~\bibnamefont {Nicastro}}, \bibinfo {author} {\bibfnamefont {R.~B.}\ \bibnamefont {Meyer}},\ and\ \bibinfo {author} {\bibfnamefont {Z.}~\bibnamefont {Dogic}},\ }\bibfield  {title} {\bibinfo {title} {Reconfigurable self-assembly through chiral control of interfacial tension},\ }\bibfield  {journal} {\bibinfo  {journal} {Nature}\ }\textbf {\bibinfo {volume} {481}},\ \href
  {https://doi.org/10.1038/nature10769} {10.1038/nature10769} (\bibinfo {year} {2012}{\natexlab{b}})\BibitemShut {NoStop}%
\bibitem [{\citenamefont {Jia}\ \emph {et~al.}(2017)\citenamefont {Jia}, \citenamefont {Zakhary}, \citenamefont {Dogic}, \citenamefont {Pelcovits},\ and\ \citenamefont {Powers}}]{Jia2017}%
  \BibitemOpen
  \bibfield  {author} {\bibinfo {author} {\bibfnamefont {L.~L.}\ \bibnamefont {Jia}}, \bibinfo {author} {\bibfnamefont {M.~J.}\ \bibnamefont {Zakhary}}, \bibinfo {author} {\bibfnamefont {Z.}~\bibnamefont {Dogic}}, \bibinfo {author} {\bibfnamefont {R.~A.}\ \bibnamefont {Pelcovits}},\ and\ \bibinfo {author} {\bibfnamefont {T.~R.}\ \bibnamefont {Powers}},\ }\bibfield  {title} {\bibinfo {title} {Chiral edge fluctuations of colloidal membranes},\ }\bibfield  {journal} {\bibinfo  {journal} {Physical Review E}\ }\textbf {\bibinfo {volume} {95}},\ \href {https://doi.org/10.1103/PhysRevE.95.060701} {10.1103/PhysRevE.95.060701} (\bibinfo {year} {2017})\BibitemShut {NoStop}%
\bibitem [{\citenamefont {Gibaud}\ \emph {et~al.}(2017{\natexlab{b}})\citenamefont {Gibaud}, \citenamefont {Kaplan}, \citenamefont {Sharma}, \citenamefont {Zakhary}, \citenamefont {Ward}, \citenamefont {Oldenbourg}, \citenamefont {Meyer}, \citenamefont {Randall D.~Kamien},\ and\ \citenamefont {Dogic}}]{Gibaud2017}%
  \BibitemOpen
  \bibfield  {author} {\bibinfo {author} {\bibfnamefont {T.}~\bibnamefont {Gibaud}}, \bibinfo {author} {\bibfnamefont {C.~N.}\ \bibnamefont {Kaplan}}, \bibinfo {author} {\bibfnamefont {P.}~\bibnamefont {Sharma}}, \bibinfo {author} {\bibfnamefont {M.~J.}\ \bibnamefont {Zakhary}}, \bibinfo {author} {\bibfnamefont {A.}~\bibnamefont {Ward}}, \bibinfo {author} {\bibfnamefont {R.}~\bibnamefont {Oldenbourg}}, \bibinfo {author} {\bibfnamefont {R.~B.}\ \bibnamefont {Meyer}}, \bibinfo {author} {\bibfnamefont {T.~R.~P.}\ \bibnamefont {Randall D.~Kamien}},\ and\ \bibinfo {author} {\bibfnamefont {Z.}~\bibnamefont {Dogic}},\ }\bibfield  {title} {\bibinfo {title} {Achiral symmetry breaking and positive {Gaussian} modulus lead to scalloped colloidal membranes.},\ }\bibfield  {journal} {\bibinfo  {journal} {Proceedings of the National Academy of Sciences}\ }\textbf {\bibinfo {volume} {114}},\ \href {https://doi.org/10.1073/pnas.1617043114} {10.1073/pnas.1617043114} (\bibinfo {year} {2017}{\natexlab{b}})\BibitemShut {NoStop}%
\bibitem [{\citenamefont {Senti}\ \emph {et~al.}(1955)\citenamefont {Senti}, \citenamefont {Hellman}, \citenamefont {Ludwig}, \citenamefont {Babcock}, \citenamefont {Tobin}, \citenamefont {Glass},\ and\ \citenamefont {Lamberts}}]{Senti1955}%
  \BibitemOpen
  \bibfield  {author} {\bibinfo {author} {\bibfnamefont {F.~R.}\ \bibnamefont {Senti}}, \bibinfo {author} {\bibfnamefont {N.~N.}\ \bibnamefont {Hellman}}, \bibinfo {author} {\bibfnamefont {N.~H.}\ \bibnamefont {Ludwig}}, \bibinfo {author} {\bibfnamefont {G.~E.}\ \bibnamefont {Babcock}}, \bibinfo {author} {\bibfnamefont {R.}~\bibnamefont {Tobin}}, \bibinfo {author} {\bibfnamefont {C.~A.}\ \bibnamefont {Glass}},\ and\ \bibinfo {author} {\bibfnamefont {B.~L.}\ \bibnamefont {Lamberts}},\ }\bibfield  {title} {\bibinfo {title} {Viscosity, sedimentation, and light‐scattering properties of fraction of an acid‐hydrolyzed dextran.},\ }\href@noop {} {\bibfield  {journal} {\bibinfo  {journal} {Journal of polymer science}\ }\textbf {\bibinfo {volume} {17}},\ \bibinfo {pages} {527} (\bibinfo {year} {1955})}\BibitemShut {NoStop}%
\bibitem [{\citenamefont {Jülicher}\ and\ \citenamefont {Seifert}(1994)}]{Julicher1994}%
  \BibitemOpen
  \bibfield  {author} {\bibinfo {author} {\bibfnamefont {F.}~\bibnamefont {Jülicher}}\ and\ \bibinfo {author} {\bibfnamefont {U.}~\bibnamefont {Seifert}},\ }\bibfield  {title} {\bibinfo {title} {Shape equations for axisymmetric vesicles: A clarification.},\ }\bibfield  {journal} {\bibinfo  {journal} {Physical Review E}\ }\textbf {\bibinfo {volume} {49}},\ \href {https://doi.org/10.1103/physreve.49.4728} {10.1103/physreve.49.4728} (\bibinfo {year} {1994})\BibitemShut {NoStop}%
\end{thebibliography}%

\clearpage

\onecolumngrid

\section*{Supporting Information}

\setcounter{figure}{0}   
\renewcommand{\thefigure}{S\arabic{figure}}

\subsection{Measuring vesicle material properties}

\subsubsection{Estimating membrane density}

To estimate the areal membrane density, we measured the density of virus rod suspensions in a concentrated solution using an oscillating U-tube density meter (DMA 4100, Anton-Paar). We found that the density of the tris buffer is $\rho_{\text{tris}} = 0.9994~\text{g/cm}^3$ and the density of the virus solution at 41 mg/mL is $\rho_\mathrm{nano385}(41~\text{mg/mL}) = 1.0130~\text{g/cm}^3$. Small-angle x-ray scattering on \textit{fd-wt} membranes at 54 mg/mL dextran found membrane concentration to be $275~\text{mg/mL}$~\cite{Balchunas2019}. We then extrapolate the density difference to be
\begin{equation}
    \Delta \rho_\mathrm{nano385} (275~\text{mg/mL}) = \frac{1.0130~\text{g/cm}^3 - 0.9994~\text{g/cm}^3}{41~\text{mg/mL}} (275~\text{mg/mL}) = 0.091~\text{g/cm}^3.  
\end{equation}
So the areal density is
\begin{equation}
    \sigma = (385~\text{nm}) (\frac{1~\text{cm}}{10^7~\text{nm}}) 0.091~\text{g/cm}^3 = 3.5 * 10^{-6}~\text{g/cm}^2, 
\end{equation}
from which we find
\begin{equation}
    \sigma g = (3.5 * 10^{-6}~\text{g/cm}^2) * (980~\text{cm/s}^2) = 3.4 * 10^{-3}~\frac{\text{g}}{\text{cm s}^2} = 0.077~\frac{k_\text{B} T}{\mu \text{m}^3}.
\end{equation}

\subsubsection{Measurement of the bending modulus using the out-of-plane fluctuation spectrum}\label{appendix:OutOfPlane}

\begin{figure}[h]
	\centering
	\includegraphics[width=0.7\textwidth]{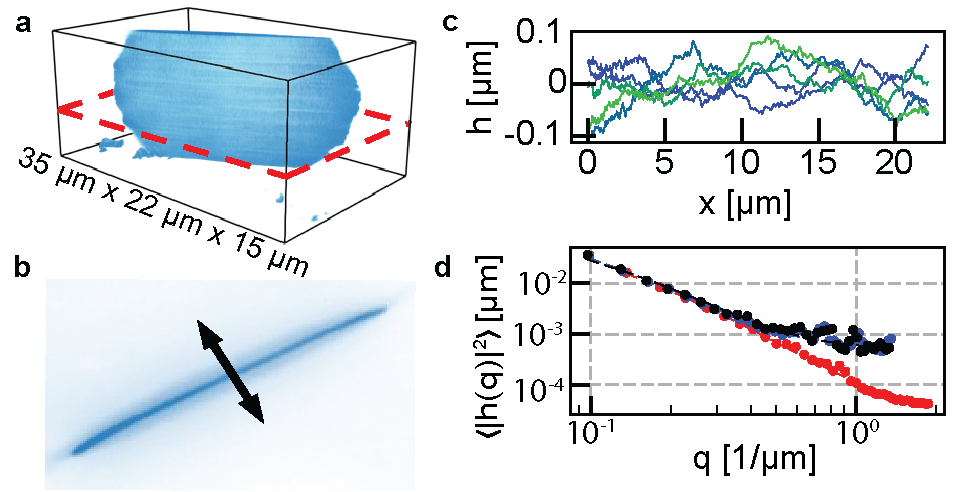}
	\caption{\textbf{Measuring bending modulus using the out-of-plane fluctuation spectrum.} \textbf{a,} Rendering of a flat nano385 membrane suspended from the chamber top. \textbf{b,} Center-slice of the membrane. \textbf{c,} The out-of-plane fluctuations were used to compute \textbf{d,} the fluctuation spectrum. Black and blue contours were taken with 100x magnification and the red contour at 150x magnification.}
	\label{fig:OutOfPlaneFluctuations}
\end{figure}

The mean bending modulus, $\kappa$, was measured by imaging the bending fluctuations of flat membranes suspended from the chamber ceiling at 20 ms intervals (Fig. \ref{fig:OutOfPlaneFluctuations}A,B). The membrane conformation was found by determining the maximum intensity along each line perpendicular to the membrane, using a first-order Savitzky-Golay filter and then refined to subpixel accuracy along each point on the contour by interpolating a 5x5 pixel neighborhood around each point along the contour. Gradients of these interpolated regions were used to find the normal and the final interface position was taken to be the point along this normal equal to a predefined intensity value. 

To find the fluctuation spectra from the membrane configuration, the mean contour was subtracted from each time point, and the signal was first multiplied by a Hanning function and rescaled to preserve the fluctuation amplitude. This procedure accounted for the non-periodicity of the membrane configuration. These processed contours were then used to calculate the power spectrum.  (Fig. \ref{fig:OutOfPlaneFluctuations}C). The power spectrum was fitted with the equation
\begin{equation}
    \langle |A^2(q) | \rangle =  \langle \epsilon^2 \rangle  + \frac{k_B T q}{\mu} \Big[1 - \frac{1}{\sqrt{1+\mu / (\kappa q^2)}} \Big]
\end{equation}
where $\langle \epsilon^2 \rangle$ is a fitting parameter for the noise and $\mu$ is the lateral tension. This equation is appropriate for fitting the fluctuations of a one-dimensional cut along a two-dimensional sheet~\cite{Mutz1990}. Measurements on three separate membranes yielded $\rho = (370 \pm 90)\,k_\text{B} T\text{/µm}^2$, $\kappa = (11000 \pm 1000)\,k_\text{B} T$ and $\langle \epsilon^2 \rangle = (3.5 \pm 0.2) *10^{-4}~\text{µm}$ (Fig. \ref{fig:OutOfPlaneFluctuations}D).  

\subsubsection{Measurement of the edge tension using the in-plane fluctuation spectrum}\label{appendix:InPlaneFluctuation}

\begin{figure}[h!]
	\centering
	\includegraphics[width=0.5\textwidth]{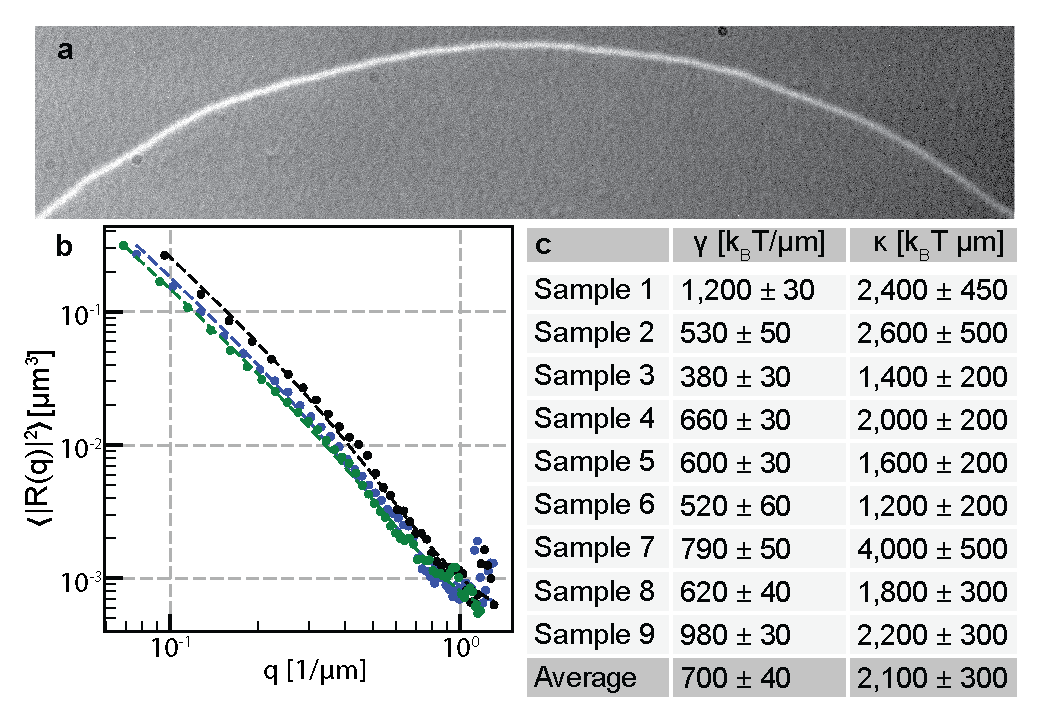}
	\caption{\textbf{In-plane fluctuation analysis to measure material properties.} (\textbf{A}) An example of a typical DIC microscopy image seen in experiment. (\textbf{B}) Fluctuation spectrum of three membranes, fitted with predicted functional form. (\textbf{C}) Table showing the measures of $\gamma$ and $\kappa_\text{B}$ for each of the nine samples analyzed. }
	\label{fig:InPlaneFluctuations}
\end{figure}

The edge tension was measured by imaging the fluctuations of flat  membranes at the chamber bottom using DIC microscopy, at 20 ms intervals (Fig. \ref{fig:InPlaneFluctuations}A)~\cite{Gibaud2012}. Edges were contoured by first finding the maximum value $h(x)$ for each point along $x$ in the image, and then refined to subpixel accuracy for each point on the contour by interpolating a 5x5 pixel neighborhood around each point $(x,y)$. Gradients of these interpolated regions were used to find the normal and the final interface position was taken to be the point along this normal equal to a predefined intensity value. The power spectrum was calculated and fitted with the equation
\begin{equation}
    \langle |B^2(q) | \rangle =  \langle \upsilon^2 \rangle  + \frac{k_B T}{\gamma q^2 + \kappa_B q^4}    
\end{equation}
where $\gamma$ is the line tension, $\langle \upsilon^2 \rangle$ is a fitting parameter for the noise and $\kappa_B$ is the edge bending energy (Fig. \ref{fig:InPlaneFluctuations}B)~\cite{Jia2017}. This was done for nine separate membranes, to measure an average of $\gamma = 700 \pm 40\,k_\text{B} T \text{/µm}$ and $\kappa_B = 2100 \pm 300\,k_\text{B} T~\text{µm}$ (Fig. \ref{fig:InPlaneFluctuations}C).

\subsubsection{Estimate of the Gaussian bending modulus}\label{Appendix:GaussianModulus}

To estimate the Gaussian bending modulus, we use the a simple argument, treating a membrane of thickness $l$ surrounded by an Asakura–Osawa ideal gas of depleting polymers~\cite{Gibaud2017}. This argument estimates the Gaussian bending modulus to be 

\begin{equation}
\Bar{\kappa} = \frac{n l^2 R_g}{6}\,k_\text{B} T 
\end{equation}

\noindent where $n$ is the depletant concentration and $R_g$ is the radius of gyration of the depleting polymer. For our nano385 membranes, $l = 385$~nm, the polymer concentration $n = 56$~mg/ml and for 500 kDa dextran the radius of gyration $R_g = 30$~nm~\cite{Senti1955}. Using these values, we find $\Bar{\kappa} = 50\,k_\text{B} T$.

\subsection{Metastability calculation}\label{Appendix:Stability}

As an initial estimate for the criteria of stability of the membrane or vesicle configurations, we calculated the energy of a membrane curved into a series of spherical caps transitioning from a flat sheet to a closed vesicle. Each of these spherical caps is a section of a sphere of radius $R$ and a height of $h$, with an area $A = 2 \pi R h$. The energy of these spherical caps is given by the Helfrich free energy~\cite{Boal1992}

\begin{equation}
E = \frac{\kappa}{2} \int (2H)^2 dA + \Bar{\kappa} \int K dA + \gamma \int dL.
\end{equation}

\noindent The bending energy simplifies to $\frac{\kappa}{2} \int (\frac{2}{R})^2 dA = \frac{2 \kappa}{R^2} A = \frac{4 \pi \kappa h }{R} $. The Gaussian modulus term become $\Bar{\kappa} \int K dA = \frac{\Bar{\kappa}}{R^2} A$ The line tension term becomes $\gamma \int dL = 2 \gamma \pi \sqrt{2 R h - h^2}$. So, the total energy of the spherical cap is 

\begin{equation}
E_\mathrm{cap} = \frac{2 \pi (2 \kappa+\Bar{\kappa}) h }{R} + 2 \gamma \pi \sqrt{2 R h - h^2}
\end{equation}
 
First, we find the area at which the energy of the membrane is equal to the energy of the vesicle, $A_1^*$. A closed vesicle occurs at $h=2R$, giving energy of

\begin{equation}
E_\mathrm{vesicle} = 4 \pi (2 \kappa + \Bar{\kappa}).
\end{equation}

\noindent The flat disk-like membrane configuration occurs at $h=0, R = \infty$, which gives an energy of

\begin{equation}
E_\mathrm{disk} = 2 \gamma \pi \sqrt{\frac{A}{\pi}}.
\end{equation}

Setting these equal,

\begin{equation}
2 \gamma \pi \sqrt{\frac{A_1^*}{\pi}} = 4 \pi (2 \kappa + \Bar{\kappa})
\end{equation}

\begin{equation}
A_1^* = \frac{4 \pi}{\gamma^2} (2 \kappa + \Bar{\kappa})^2.
\end{equation}

Next, we study the energy barrier for vesicle closure and rupture.  Since

\begin{equation}
E_\mathrm{cap} = \frac{2 \pi (2 \kappa + \Bar{\kappa}) h }{R} + 2 \gamma \pi \sqrt{2 R h - h^2} = \frac{(2 \kappa + \Bar{\kappa}) A }{ R^2} + 2 \gamma \pi \sqrt{\frac{A}{\pi}- \frac{(A)^2}{4 \pi^2 R^2 }}, 
\end{equation}

\noindent the derivative at fixed area is

\begin{equation}
\frac{\partial E_\mathrm{cap}}{\partial R}  = - \frac{2 (2 \kappa + \Bar{\kappa}) A }{ R^3} + \gamma \frac{ (A)^2}{ 2 \pi  R^3 \sqrt{\frac{A}{\pi}- \frac{(A)^2}{4 \pi^2 R^2 }}},
\end{equation}

\noindent which vanishes when,

\begin{equation}
A = \frac{4 \pi  (4 \kappa + 2\Bar{\kappa} )^2  }{ \gamma^2 + (\frac{4\kappa + 2\Bar{\kappa}}{R})^2  },
\end{equation}

\noindent or,

\begin{equation}
~\ R = R^* = \frac{\sqrt{A}}{\sqrt(4 \pi - \gamma^2 A / (4 \kappa + 2 \bar{\kappa})^2)}.
\end{equation}

\noindent The energy barrier for closure is 

\begin{equation}
E_b^\mathrm{close} = E_\mathrm{cap} - E_\mathrm{disk}|_{R = R^*} = \frac{[\gamma \sqrt{A} - 2\sqrt{\pi}/(4 \kappa + \bar{}\kappa)]^2}{4 ( \kappa + \bar{\kappa})}.
\end{equation}

\noindent Note that the barrier decreases with area, vanishing at $A = A_2^* = 4 \pi (4 \kappa + 2 \bar{\kappa})^2 / \gamma^2$. On the other hand, there is always a barrier to forming a single pore on a spherical membrane: 

\begin{equation}
E_b^\mathrm{rupture} = E_\mathrm{cap} - E_\mathrm{vesicle}|_{R = R^*} = \frac{\gamma^2 A}{4 (2 \kappa + \bar{\kappa})}.
\end{equation}

\noindent While these spherical cap calculations roughly capture the features of a membrane with a single pore, the barrier is underestimated compared to numerical solutions of the Euler-Lagrange equations~\cite{Boal1992}. We study the single- and double-pore shapes numerically in the next section.

For the experimentally measured values of $\gamma \approx 700~\frac{k_\text{B} T}{\text{µm}}$ and $\kappa \approx 11,000\,k_\text{B} T$ and $\Bar{\kappa} = 50\,k_\text{B} T$, we get $A_1^* = \frac{4 \pi}{\gamma^2}  (2 \kappa + \Bar{\kappa})^2$ = 12,500 µm$^2$ and $A_2^* = \frac{4 \pi}{\gamma^2}  (4 \kappa + 2 \Bar{\kappa})^2$ = 50,000 µm$^2$. At areas below 13,000 µm$^2$, membranes are the only energetic minima, while at areas above 49,800 µm$^2$, vesicles are the only energetic minima. For areas between these two values, the flat membrane is the global energetic minima, but vesicles are the local energetic minima and are therefore metastable.

\subsection{Numerical \textbf{methods} for shape analysis}\label{Appendix:2}

\subsubsection{Energy and coordinate system}

\begin{figure}[h!]
	\centering
	\includegraphics[width=0.3\textwidth]{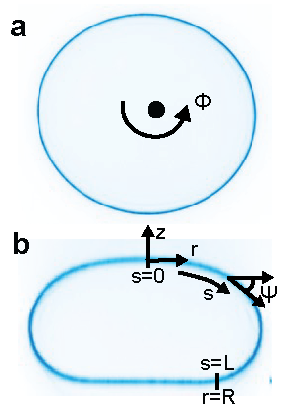}
	\caption{\textbf{Axisymmetric coordinate system for vesicle shape analysis.} \textbf{a,} Equatorial and \textbf{b,} azimuthal vesicle cross sections with overlaid coordinate system. The vesicle contacts the floor at $r=R$, and $s$ is the arclength.}
	\label{fig:ColloidosomeCrossSection}
\end{figure}

For shape analysis, we focus on axisymmetric membranes, using the coordinates arclength $s$ and azimuthal angle $\phi$, with corresponding unit vectors
\begin{equation} \label{eq:C1}
	\mathbf{e}_s = \begin{pmatrix} r'(s) \\ z'(s) \\ 0 \end{pmatrix} =
	\begin{pmatrix} \cos \psi(s) \\ -\sin \psi(s) \\ 0 \end{pmatrix}, \quad
	\mathbf{e}_\phi = \begin{pmatrix}
		-r(s) \sin(\phi) \\
		r(s) \cos(\phi) \\
		0
	\end{pmatrix},
\end{equation}
where $\psi(s)$ is the local tangent angle (Fig. 
\ref{fig:ColloidosomeCrossSection}). 
	
For the most general case, we have to consider the gravitational energy, curvature energy and edge terms
\begin{multline}\label{eq:C2}
    E = \int\! \left[\frac{\kappa}{2} \big(2 H \big)^2 + \bar{\kappa} K_G + \sigma g z + \mu \right] dA +  \int\! q (r_s - \cos \psi) ds \\
 +  \int \! \eta (z_s + \sin \psi) ds + \gamma \int \! dL + P \int\! dV
\end{multline}
with the Lagrange multipliers $\mu$ to keep the surface area constant, $q(s)$ to account for $r_s = \cos \psi$, $\eta(s)$ to account for $z_s = -\sin \psi$ and $P$ to account for the constant volume~\cite{kraus1995gravity}. In this parameterization, the energy can be written as
\begin{multline}\label{eq:C3}
E = 2 \pi \int \left[\frac{\kappa}{2} \big( \psi_s + \frac{\sin \psi}{r}  \big)^2  +\overline{\kappa} \psi_s \frac{\sin \psi}{r} + \sigma g z + \mu  \right] r ds + 2 \pi \int q (r_s - \cos \psi) ds \\
+ 2 \pi \int \eta (z_s + \sin \psi) ds + \gamma \int dL + P \frac{\pi}{2} \int r^2 \sin \psi ds
\end{multline}

\noindent From this general case, we  derive shape equations for experimentally relevant examples, by converting into a system of solvable differential equations~\cite{Julicher1994}.

\subsubsection{Closed colloidosomes}
	
For closed vesicles, the line tension term is zero, so that the total energy is
\begin{multline}\label{eq:C4}
    E = 2 \pi \int \Big[\frac{\kappa}{2} \big( \psi_s + \frac{\sin \psi}{r}  \big)^2  +\overline{\kappa} \psi_s \frac{\sin \psi}{r} + \sigma g z + \mu  \Big] r ds + 2 \pi \int q (r_s - \cos \psi) ds \\ 
    + 2 \pi \int \eta (z_s + \sin \psi) ds+ P \frac{\pi}{2} \int r^2 \sin \psi ds.
\end{multline}
First, we take the variation along $\psi$ giving	
\begin{equation}
    - \kappa r \psi_{ss} - \kappa \cos \psi \psi_s + \kappa \frac{\sin \psi \cos \psi}{r} + q \sin \psi + \eta \cos \psi + \frac{P}{2} r^2 \cos \psi  = 0
\end{equation}
The variation along $r$ gives
\begin{equation}
    \frac{\kappa}{2} \psi_s^2 - \frac{\kappa}{2} \frac{\sin^2 \psi}{r^2} + \mu - q_s + \sigma g z + P r \sin \psi = 0.    
\end{equation}
Taking the variation with respect to $z$ gives
\begin{equation}
   \eta_s = \sigma g r.
\end{equation}
	
We can then define the arclength in terms of the reduced arclength so that $T = s/L$, where $L$ is the total length of the curve. The above equations, along with the constraints that $\mu$, $L$ and $P$ are constant, can be recast into a system of first-order ODEs:
\begin{subequations} \label{eq:VesicleEquations}
\begin{align}
    \frac{d \psi}{dT} & = L \psi_s \\
    \frac{d \psi_s}{dT} &=  - L \frac{\cos \psi \psi_s}{r}  +  L \frac{\sin \psi \cos \psi}{r^2} + L \frac{q}{r \kappa} \sin \psi + L \frac{\eta}{r \kappa} \cos \psi + L \frac{P}{2} r^2 \cos \psi \\
    \frac{d r}{d T} &= L \cos \psi\\
    \frac{d z}{d T} &= - L \sin \psi \\
    \frac{d q}{d T} &= L \frac{\kappa}{2} \psi_s^2 - L \frac{\kappa}{2} \frac{\sin^2 \psi}{r^2} + L \mu + L \sigma g z + L P r \sin \psi \\
    \frac{d A}{d T} &= 2 \pi r L \\
    \frac{d \mu}{d T} &= 0 \\
    \frac{d \eta}{d T} &= L \sigma g r\\
    \frac{d L}{d T} &= 0\\
    \frac{d P}{d T} &= 0\\
    \frac{d V}{d T} &= L r^2 \sin \psi.
\end{align}
\end{subequations}

We need eleven boundary conditions to solve this system. To begin with, we enforce five boundary conditions at the top center of the vesicle where the tangent angle is zero, the radius is zero, the height is zero, the integrated area is zero and the integrated volume is zero, leading to the five conditions:
\begin{equation} \label{eq:VesicleBoundary1}
    \psi(0) = 0, r(0) = 0, z(0) = 0, A(0) = 0, V(0) = 0.
\end{equation}	
At the bottom of the vesicle, the radius in contact with the floor is set to $R$ (Fig.~\ref{fig:ColloidosomeCrossSection}). The integrated area is $A_i - \pi R^2$, the integrated volume is $V_i$ and the tangent angle is $\pi$, giving the four conditions:
\begin{equation}\label{eq:VesicleBoundary2}
    r(T=1) = R, A(1) = A_i - \pi R^2, \psi(1) = \pi, V(1) = V_i
\end{equation}
We also have two additional conditions. The first is that the free energy is zero at the top by the transversality condition. This gives
\begin{align}
    \frac{\mathcal{H}(0)}{2 \pi} &= \psi_s \frac{\partial \mathcal{L}}{\partial \psi_s} + r_s \frac{\partial \mathcal{L}}{\partial r_s} - \mathcal{L} = 0 \\
    &= \frac{\kappa r}{2} (\psi_s^2 - \sin^2(\psi)/r^2) - \mu r + q \cos(\psi) - \eta  \sin(\psi) - \sigma g r z |_{s = 0}
\end{align}
which reduces to 
\begin{equation}\label{eq:VesicleBoundary3}
    q(0) = 0.
\end{equation}
The final boundary condition comes from having zero imposed force at the top, 
\begin{equation} \label{eq:final-bc}
    \eta(0) = 0.
\end{equation}
This system of equations \eqref{eq:VesicleEquations}--\eqref{eq:final-bc} defines a boundary value problem that we solved numerically using scipy function \textit{solve\_bvp()}. The control parameters are the radius of the contact with the bottom surface, the total surface area, and the total volume as free parameters, which can each be experimentally measured. To determine the volume at zero osmotic pressure difference, $V_0$, we numerically solve for the shape at $P=0$, which we then use to find $\lambda = V / V_0$.

\subsubsection{Pendent colloidosomes}
\label{appendix:PendentVesicle}

The main difference between modeling pendent vesicles and closed vesicles is the presence of an open pore at the top. However, since the width of the top opening is a fixed radius measured experimentally, the edge tension term does not change under variation of the shape. Therefore, we may use the same energy, Eq.~\ref{eq:C4}, and the Euler-Lagrange equationsfor pendent colloidosomes are be identical to Eq.~\ref{eq:VesicleEquations}. However, the boundary conditions are changed.
	
To begin with, we enforce the four boundary conditions at the top of the pendent membrane $z = 0$: the radius is zero, the integrated area is zero, the integrated volume  is zero, and the tangent angle is zero,
\begin{equation}
    z(0) = 0, r(0) = 0, A(0) = 0, V(0) = 0, \psi(0) = 0.
\end{equation}
At the bottom of the pendent vesicle 
the radius is fixed to be a constant $R$, the integrated area is $A_i$, the integrated volume is $V_i$, and the bending moment at the interface must vanish, giving the five conditions
\begin{equation}
     r(1) = R, A(1) = A_i, V(1) = V_i, \psi_s(1) =  - \frac{\sin \psi (1)}{r(1)}.
\end{equation}
Finally, the conditions $q(0) = 0$ and $\eta(0) = 0$ are the same as for the closed vesicle.
We solve these equations with experimentally measured values for the radius of the top opening, $R$, the integrated area and the enclosed volume. To determine the volume at zero osmotic pressure difference, $V_0$, we numerically solve for the shape at $P=0$, which we then use to find $\lambda = V / V_0$.

\subsubsection{Single pore colloidosomes}
\label{appendix:OnePore}
	
To model the single pore evaporating vesicles, we use the energy functional in Eq.~\ref{eq:C4}, with minor modifications. First, since the vesicle has an open pore, the volume is no longer fixed and instead the pressure is equilibrated at $P = 0$. We also assume that the gravitational energy does not significantly contribute, since the vesicles we observe are small. The boundary value problem is therefore equivalent to the pendent case with $g = 0$. This reduces the set of ODEs in Eq.~\ref{eq:VesicleEquations} to

\begin{subequations} 
\label{eq:SinglePoreEquations}
\begin{align}
    \frac{d \psi}{dT} & = L \psi_s \\
    \frac{d \psi_s}{dT} &=  - L \frac{\cos \psi \psi_s}{r}  +  L \frac{\sin \psi \cos \psi}{r^2} + L \frac{q}{r \kappa} \sin \psi + L \frac{\eta}{r \kappa} \cos \psi  \\
    \frac{d r}{d T} &= L \cos \psi\\
    \frac{d z}{d T} &= - L \sin \psi \\
    \frac{d q}{d T} &= L \frac{\kappa}{2} \psi_s^2 - L \frac{\kappa}{2} \frac{\sin^2 \psi}{r^2} + L \mu\\
    \frac{d A}{d T} &= 2 \pi r L \\
    \frac{d \mu}{d T} &= 0 \\
    \frac{d \eta}{d T} &= 0\\
    \frac{d L}{d T} &= 0\\
\end{align}
\end{subequations}

\noindent To solve, we require nine boundary conditions. We start with four boundary conditions at the top of the membrane, $z = 0$: the radius is zero, the integrated area is zero, and the tangent angle is zero,

\begin{equation}
    z(0) = 0, r(0) = 0, A(0) = 0, \psi(0) = 0.
\end{equation}

\noindent We also have the three boundary conditions at the bottom of the membrane: the radius of the pore is $r_1$, the integrated area is $A_i$ and the bending moment vanishes at the edge of the pore. Together these are

\begin{equation}
     r(1) = r_1, A(1) = A_i, \psi_s(1) =  - \frac{\sin \psi (1)}{r(1)}.
\end{equation}

\noindent Finally, the conditions $q(0) = 0$ and $\eta(0) = 0$ are the same as for the closed vesicle. It is important to note that since $g = 0$, the equations are translationally invariant along $z$ so that vanishing force at one boundary ($\eta(T = 0) = 0$) automatically implies that it also vanishes at the other boundary ($\eta(T = 1) = 0$). In fact $\eta(T) = 0$ identically.

\subsubsection{Two pore colloidosomes}\label{appendix:TwoPore}

To model the two-pore system, we use the same ODEs as the single pore membrane, Eq.~\ref{eq:SinglePoreEquations}. To solve, we again require nine boundary conditions. We start with four boundary conditions at the top of the membrane $z = 0$: the radius of the top pore is $r_1$, the integrated area is zero, and the bending moment vanishes at the edge of the pore,

\begin{equation}
    z(0) = 0, r(0) = r_1, A(0) = 0, \psi_s(0) =  - \frac{\sin \psi (0)}{r(0)}.
\end{equation}

\noindent We also have the three boundary conditions at the bottom of the membrane: the radius of the bottom pore is $r_2$, the integrated area is $A_i$ and bending moment again vanishes at the edge of the pore. Together, these are

\begin{equation}
     r(1) = r_2, A(1) = A_i, \psi_s(1) =  - \frac{\sin \psi (1)}{r(1)}.
\end{equation}

\noindent Finally, the conditions $q(0) = 0$ and $\eta(0) = 0$ are the same as for the closed vesicle.

\clearpage

\begin{figure}[h]
	\centering
	\includegraphics[width=0.5\textwidth]{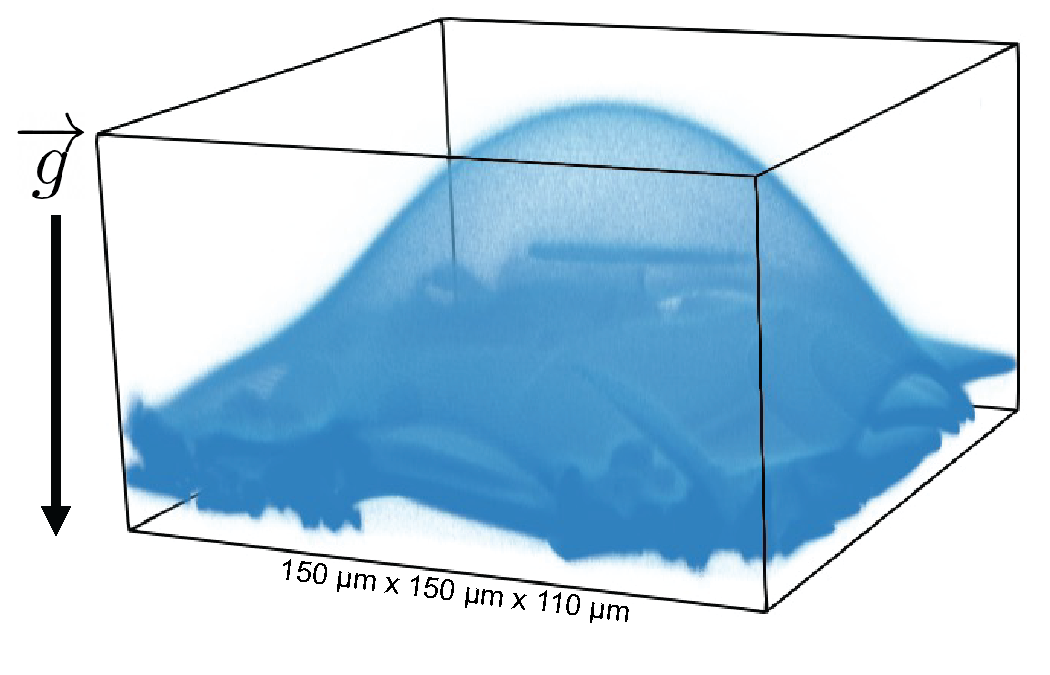}
	\caption{\textbf{Membrane prior to chamber inversion.} Typical membrane seen at the bottom of the chamber, just before inversion. The membrane is curving upward but has not closed into a vesicle.}
	\label{fig:CurvingVesicle}
\end{figure}

\begin{figure}[h]
	\centering
	\includegraphics[width=0.6\textwidth]{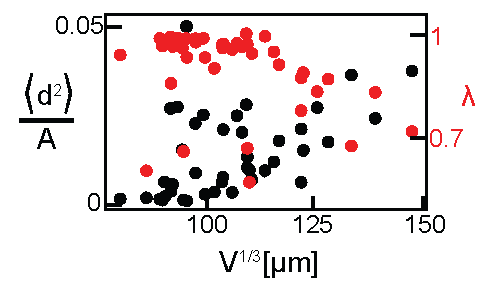}
	\caption{\textbf{Evaluating fit and inflation against vesicle volumes.} Value of $\frac{<d^2>}{A}$ (black) and reduced volume $\lambda$ (red) for N=42 nano385 colloidosomes.}
	\label{fig:ColloidosomeFitAndLambda}
\end{figure}

\begin{figure}[h]
	\centering
	\includegraphics[width=0.6\textwidth]{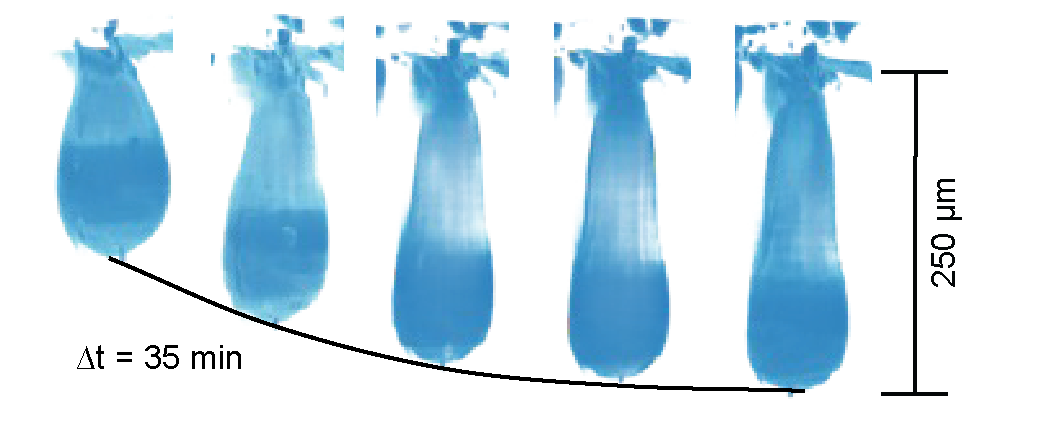}
	\caption{\textbf{Pendent membrane that remains suspended from the top surface.} Example of a pendent membrane which extends for roughly 1 hour after inversion, before stabilizing at a finite length.}
	\label{fig:SecondPendentVesicle}
\end{figure}

\begin{figure}[h]
	\centering
	\includegraphics[width=0.8\textwidth]{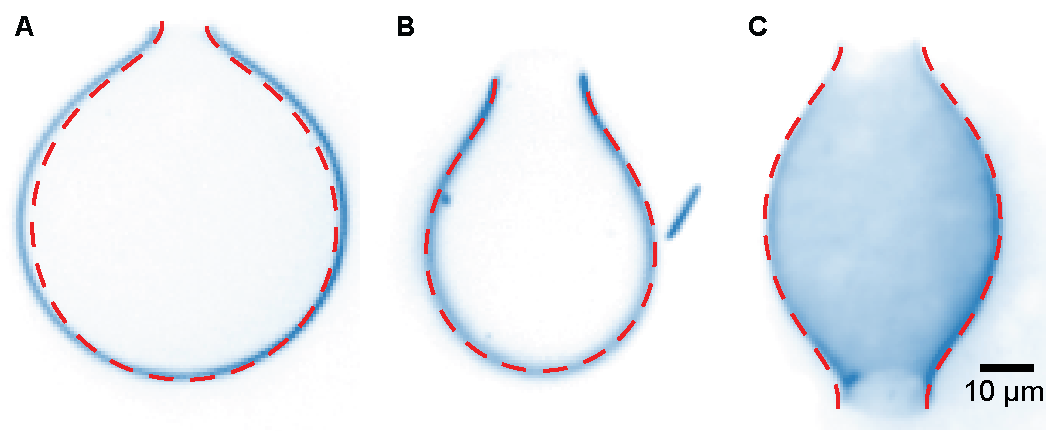}
	\caption{\textbf{Predicted one and two pore membrane shapes plotted over experimental data.} Membrane shape (\textbf{A}) one minute and (\textbf{B}) 15 minutes after a single pore nucleates. (\textbf{C})  Two-pore contour over experimental data. All predicted shapes are in good agreement with the experimentally-observed shapes.}
	\label{fig:FittedVesicles}
\end{figure}

\begin{figure}[h!]
	\centering
	\includegraphics[width=0.8\textwidth]{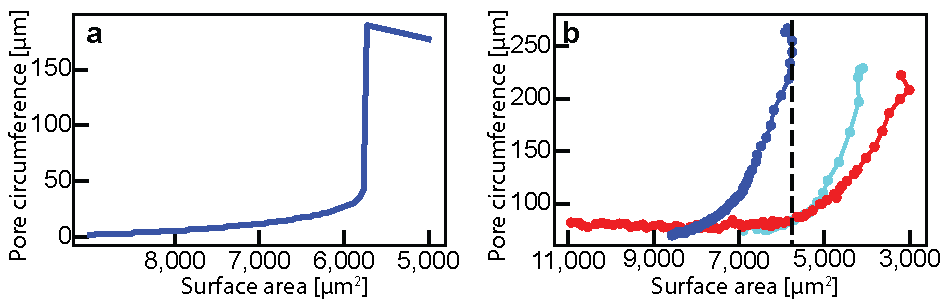}
	\caption{\textbf{Pore size during single-pore disassembly.} (\textbf{A}) Radius of pore for an energy-minimizing membrane increases as the area decreases. At a critical area of $\approx$ 5,800 µm$^2$, the energy-minimizing shape becomes a flat membrane, leading to a sharp rise in the pore radius.  (\textbf{B}) Pore circumference measured in three experimental samples. The pore circumference rises sharply. The predicted critical value of 5,800 µm$^2$ is marked by a dashed line.}
	\label{fig:VesicleInstability}
\end{figure}

\begin{figure}[h!]
	\centering
	\includegraphics[width=0.5\textwidth]{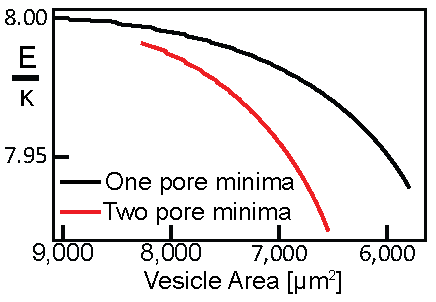}
	\caption{\textbf{Comparing the one- and two-pore energy minima.} Comparison between the one pore (black) and two pore (red) energy minima over their ranges of stability. For all areas where the two pore membranes are stable, they have a lower energy minima than the one pore membranes.}
	\label{fig:1vs2poreEnergy}
\end{figure}

\clearpage

\noindent\textbf{Movie S1.} Many vesicles form in high-purity samples. Following our purification procedure, hundreds of vesicles can be assembled at once. In this field of view (2.3 mm x 2 mm x 150 µm), there are $\approx$80 vesicles of variable sizes and shapes.

\noindent\textbf{Movie S2.} Pendent colloidosome tearing and vesicle closure. To form closed vesicles, a pendent colloidosome tears near the top surface, leaving a vesicle with a single pore held open by a tether. Once this tether breaks, the pore seals, resulting in a closed vesicle. This movie shows two such examples of this tearing and closure process.

\noindent\textbf{Movie S3.} Vesicles follow a similar dynamic pathway during disassembly. Four vesicles simultaneously disassembling according to the one-pore pathway. These vesicles all decrease in area, and follow a nearly identical pore-opening process once the surface area reaches a critical value.

\noindent\textbf{Movie S4.} Single-pore colloidosome disassembly. A vesicle unwrapping according to the single-pore disassembly pathway. The first clip shows an example of this process in experiment. The second clip shows the pathway predicted by energy minimization.

\noindent\textbf{Movie S5.} Transient pore formation on vesicle surface.  Vesicles which disassemble slowly begin to have transient pores that open and close on their surface. One example of this behavior is shown from two orthogonal views.

\noindent\textbf{Movie S6.} Two-pore colloidosome disassembly. A vesicle unwrapping according to the two-pore disassembly pathway. The first movie segment shows an example of this process in experiment. This vesicle is shown from above (left column) and from an angled viewpoint (right column) with both the raw fluorescent images (top row) and the processed meshes (bottom row). The second movie segment shows the pathway predicted by energy minimization.

\end{document}